\documentclass[final,3p,times,authoryear,12pt]{elsarticle}
\usepackage{amsmath,amssymb,amsfonts}
\usepackage{import}
\usepackage{array,multirow}
\usepackage[table]{xcolor}
\usepackage{fancyhdr}
\usepackage{natbib}
\usepackage{graphicx}
\usepackage{pdflscape}
\usepackage{color}
\usepackage{tabularx,ragged2e,booktabs}
\usepackage{graphics}
\usepackage{url} 
\usepackage{wrapfig}
\usepackage{lscape}
\usepackage{rotating}
\usepackage{epstopdf}
\usepackage{caption}
\usepackage{mathtools}
\usepackage{fancyvrb}
\usepackage{subcaption}
\usepackage{fixltx2e} 
\usepackage{flexisym}

\newcommand{\der}[2]{\frac{\partial #1 }{\partial #2}}

\newcommand\val[1]{v_{#1}}

\newcommand\es[1]{e_{#1}}

\newcommand\valfor[1]{\hat{v}_{#1}}
\newcommand\esfor[1]{\hat{e}_{#1}}

\newcommand\E{\operatorname{E}}
\newcommand{\Var}{\mathop{\mbox{\sf Var}}}

\newcommand\Sk{\operatorname{Sk}}
\newcommand\Ku{\operatorname{Ku}}
\newcommand\RV{\operatorname{RV}}

\newcommand{\bz}{\mathbf{z}}
\newcommand{\bg}{\mathbf{g}}
\newcommand{\btheta}{\boldsymbol{\theta}}
\newcommand{\bgamma}{\boldsymbol{\gamma}}
\newcommand{\hbtheta}{\hat{\btheta}}
\newcommand{\hparv}{\hat\btheta^{(v)}}

\newcommand{\mft}{\mathfrak{t}}

\newcommand{\hbgamma}{\hat{\bgamma}}
\newcommand{\bX}{\mathbf{X}}
\newcommand{\tbX}{\widetilde{\bX}}
\newcommand{\bM}{\mathbf{M}}
\newcommand{\bH}{\mathbf{H}}
\newcommand{\hbM}{\widehat{\bM}}
\newcommand{\bHit}{\mathbf{Hit}}
\newcommand{\hbHit}{\widehat{\bHit}}
\newcommand{\tbHit}{\widetilde{\bHit}}
\newcommand{\hHit}{\widehat{Hit}}
\newcommand{\tHit}{\widetilde{Hit}}

\newcommand\parv{\btheta^{(v)}}
\newcommand\parj{\btheta^{(j)}}
\newcommand\estiv{\hbtheta^{(v)}}
\newcommand\estij{\hbtheta^{(j)}}

\definecolor{darkgreen}{rgb}{0.0, 0.7, 0.0}

\newcommand{\VaR}{$\mathsf{VaR}$}
\newcommand{\ES}{$\mathsf{ES}$}

\usepackage{hyperref}
\hypersetup{
	colorlinks,
	linkcolor = {red!80!black},
	citecolor = {blue!80!black},
	urlcolor = {red!80!black},
}

\journal{Quantitative Finance}

\begin{document}

\begin{frontmatter}

\title{Dynamic tail risk forecasting: what do realized skewness and kurtosis add?}

\author[adrG1]{Giampiero M. Gallo\fnref{fn1}}
\ead{giampiero.gallo@nyu.edu}

\author[adrO1,adrO2]{Ostap Okhrin}
\ead{ostap.okhrin@tu-dresden.de}

\author[adrS1]{Giuseppe Storti\corref{cor1}\fnref{fn2}}
\ead{storti@unisa.it}

\fntext[fn2]{Opinions expressed here are personal and do not involve the Corte dei conti.}
 
 \cortext[cor1]{Corresponding author}

 \address[adrG1]{Corte dei conti, New York University in Florence, and CRENoS}
 
  \address[adrO1]{Technische Universität Dresden, 01062 Dresden, Germany}
  \address[adrO2]{Center for Scalable Data Analytics and Artificial Intelligence (ScaDS.AI) Dresden/Leipzig, Germany}

 \address[adrS1]{Universit\`a di Salerno, Department of Economics and Statistics, Fisciano, Italy}

\begin{abstract}
\noindent This paper compares the accuracy of tail risk forecasts with a focus on including realized skewness and kurtosis in "additive" and "multiplicative" models. Utilizing a panel of 960 US stocks, we conduct diagnostic tests, employ scoring functions, and implement rolling window forecasting to evaluate the performance of Value at Risk (VaR) and Expected Shortfall (ES) forecasts. Additionally, we examine the impact of the window length on forecast accuracy. We propose model specifications that incorporate realized skewness and kurtosis for enhanced precision. Our findings provide insights into the importance of considering skewness and kurtosis in tail risk modeling, contributing to the existing literature and offering practical implications for risk practitioners and researchers.
\vspace{.6cm}
\end{abstract}


\end{frontmatter}

\section{Introduction}

Starting approximately thirty years ago, the issue of capital adequacy has received increased attention, with significant impetus given to supervisory and regulatory functions to closely monitor the impact of volatility and interconnectedness on financial institution portfolios. Modern risk management is based on the principle that increased risks must be adequately covered with sufficient resources to avoid liquidity crises or defaults that could affect other institutions and the financial system as a whole. The consequences of the 2007-2008 financial crisis underscored the need for suitable capital risk measures exhibiting forecastability over relevant time horizons.

The various recommendations of the Basel Committee on Banking Supervision regarding capital risk regulations emphasize that the main parameters of a conditional distribution of returns to be monitored are some position index, specifically the threshold \citep[Value at Risk, \VaR, ][]{jorion1997value} corresponding to a certain probability in the tail where losses occur, and the average value of the loss once that threshold has been surpassed \citep[Expected Shortfall, \ES, ][]{artzner1999}. In this context, without loss of generality, we assume that the tail in question is the left tail, representing losses in long positions.

Market activity, characterized by price and volume movements in response to news, renders the conditional distribution of returns non-constant over time. Consequently, both Value at Risk (\VaR) and Expected Shortfall (\ES) become time-varying risk measures. Moreover, observed persistence in market behavior suggests dynamics that leverage valuable past information. From an econometric perspective, it is challenging to determine which features of past market behavior are relevant for predicting \VaR\ and \ES, as these measures represent conditional quantiles and expectations, respectively, in the tail of the asset return distribution. 

Approaches to address this issue can broadly be categorized into three main groups. The first category assumes a known parametric distribution for returns, typically a Student-$t$ distribution, and focuses on the dynamic evolution of the conditional variance of returns. This approach augments the fixed quantile identification with a GARCH process that models the dependence of conditional variance on recent returns and past estimates. Parameters are estimated using (Quasi) Maximum Likelihood (QML) methods. 
At the opposite end of the spectrum, parametric assumptions about the return distribution or its dynamics are entirely discarded. So-called historical simulation methods are employed, where future outcomes are simulated by repeating observed past behaviors. 

A third stream adopts an intermediate stance, focusing on the dynamics of the risk measure of interest while limiting or avoiding reliance on parametric assumptions about the shape of the conditional return distribution. This semi-parametric approach to financial risk modeling is gaining popularity due to its flexibility and often demonstrates competitive performance compared to more complex parametric models. In what follows, we will position ourselves in this stream of literature, addressing, in particular, the role that higher-order conditional moments, notably skewness and kurtosis, have on the refinement of predictions, hence highlighting the role of the time-varying evolution of asymmetry and tail density of the return distribution in sharpening the projections of \VaR\ and \ES.

Our synthesis in this field is to identify two main categories of semi-parametric modeling approaches for tail risk measures. The first is called the ``additive'' approach, which utilizes linearized representations of GARCH models, such as in CAViaR models. The second approach, referred to as the ``multiplicative'' approach, involves estimating GARCH-type models via the minimization of a properly defined strictly consistent scoring function. We consider the recent literature \citep[e.g.,][]{neuberger2013,neuberger2020,baelee2020} on the derivation of realized measures of skewness and kurtosis as consistent estimates of the conditional skewness and kurtosis of daily returns. For our purposes, these additional features of the conditional distributions may be relevant when included in the specifications. Given that \cite{amaya2015} provides evidence that realized skewness and kurtosis are useful when forecasting the cross-section distribution of equity returns, our interest here is to assess whether these benefits extend to risk forecasting as well.

Although the additive approach has gained popularity, there is still a lack of extensive forecasting comparison between these two methodologies. Hence, we aim to bridge a gap in the literature by proposing an application that evaluates the accuracy of forecasts generated by additive and multiplicative modeling strategies for a panel of 960 US stocks. To achieve this, we employ various diagnostic tests and scoring functions for both \VaR\ and \ES\ forecasts. Additionally, we investigate the impact of window length on forecasting accuracy, a critical issue for practitioners. Short windows tend to minimize bias but increase variability in risk forecasts, while long windows have the opposite effect: hence, we conduct a rolling window estimation/forecasting exercise and evaluate the performance of three different window lengths,  500, 1000, and 2000 days.

 Our novel model specifications using information on realized higher-order moments to forecast tail risk measures are both regression quantile time series models for forecasting \VaR, as well as bivariate semi-parametric models for joint \VaR\ and \ES\ forecasting: we are interested in providing specific evidence on the relevance of the realized skewness and kurtosis via Wald-type tests, but also on their contribution in improving the forecast performance, assessed with standard backtesting procedures. Their predictive performances are compared to those of competitors that do not include such information.

In a nutshell, the evidence on the vast panel of stock indicates that multiplicative models are preferred to additive ones, and that the extension to higher moments does not buy a generalized relevant improvement in the outcome. In general, simpler models are to be preferred to more complex ones.

The structure of the paper is as follows. In Section \ref{s:mods}, we propose our models in Subsection \ref{s:setup}, while the related estimation procedures and some properties of the estimators are illustrated in Subsection \ref{s:esti}. In Section \ref{s:revsk}, we present a recent literature review on realized estimators of skewness and kurtosis of financial returns. Section \ref{s:appli} is dedicated to the empirical application, while Section \ref{s:conc} concludes.

\section{Semi-parametric risk modeling}
\label{s:mods}
The literature on semi-parametric risk modeling features a seminal paper by \cite{caviar}, who introduced the Conditional Autoregressive Value-at-Risk (CAViaR) model for forecasting \VaR. This model has interesting connections with both quantile regression and GARCH models, in that the CAViaR model can be viewed as a quantile autoregression with a recursive term. By the same token, a linear GARCH model of a given order can be represented as a CAViaR model of the same order. Building on the duality between GARCH and CAViaR, \cite{Xiao_Koenker_2009} present an original approach to estimating parameters of a GARCH model, proposing to minimize the typical quantile loss function used in quantile regression models.

Direct semi-parametric modeling of \ES\ is not feasible because, unlike \VaR, \ES\ is not elicitable relative to a given loss function. However, \cite{fissleretal2015} have derived a class of loss functions that are strictly consistent for the pair (\VaR, \ES), in the sense that the expected loss is minimized by the true (\VaR, \ES). Within this framework \cite{tayl2019} proposes a class of semi-parametric models for (\VaR, \ES), augmenting the standard CAViaR setup with an additional dynamic equation for \ES, and replacing the usual quantile loss with a member of the Fissler-Ziegel (FZ) class. In particular, among the available choices, \cite{taylor2017} considers a loss, or scoring, function based on the Asymmetric Laplace quasi-likelihood function, AL for short. \cite{pattonetal2019} extend the work by \cite{tayl2019} in two different directions. First, they consider time-varying semi-parametric (\VaR, \ES) models based on the Generalized Autoregressive Score (GAS) framework \citep{crealetal2013}. Second, as done by \cite{Xiao_Koenker_2009} for \VaR, they consider directly estimating GARCH models minimizing a specific strictly consistent loss function in the FZ class called FZ0 (owing its denomination to the fact that, when using this loss to compare two models, it yields loss differentials that are homogeneous of degree zero). This property can lead to a higher power in Diebold-Mariano tests \citep{diebold1995comparing}.

\subsection{The model setup}
\label{s:setup}
Let $r_t$ be the log-return for the day $t$, for $t=1,\ldots, T$, and  $Q_{\alpha,t} = F_{r}^{-1}(\alpha|\mathcal{I}_{t-1})$ indicate the conditional $\alpha$-quantile of $r_t$ (level-$\alpha$ Value-at-Risk --\VaR), with $F_r$ being the cdf of $r_t$; correspondingly, $ES_{\alpha,t}= \E(r_{t}|r_{t}<Q_{\alpha,t},\mathcal{I}_{t-1})$ indicates the conditional $\alpha$-tail expectation of $r_t$, given past information $\mathcal{I}_{t-1}$ (level-$\alpha$ Expected Shortfall -- \ES).   

We let $\RV_t$, $\Sk_t$, and $\Ku_t$ denote, respectively, the conditional variance, skewness, and kurtosis of daily returns $r_t$ as follows 
\begin{align*}
	\RV_{t} &= \E_0\{(r_t-\mu_{1t})^2|\mathcal{I}_{t-1}\} = \mu_{2t} - \mu_{1t}^2 ,\\
	\Sk_{t} &= \E_0 \left\{\left(\frac{r_t-\mu_{1t}}{\RV^{1/2}_t}\right)^3| \mathcal{I}_{t-1}\right\}= \frac{\mu_{3t} - \mu_{1t}\mu_{2t}+2\mu_{1t}^3}{(\mu_{2t} - \mu_{1t}^2)^{3/2}} ,\\
	\Ku_{t} &= \E_0 \left\{\left(\frac{r_t-\mu_{1t}}{\RV^{1/2}_t}\right)^4| \mathcal{I}_{t-1}\right\}=\frac{\mu_{4t} - 4\mu_{3t}\mu_{1t}+6\mu_{2t}\mu_{1t}^2-3\mu_{1t}^4}{(\mu_{2t} - \mu_{1t}^2)^{2}},
\end{align*}
\noindent where $\mu_{kt} = \E_0(r_t^k|\mathcal{I}_{t-1})$ indicates the $k$-th conditional noncentered moment under the true measure. Estimates of these quantities can be readily obtained by replacing the involved conditional moments $\mu_{kt}$ with their estimated counterparts, at least using daily observations. In Section \ref{s:revsk}, we will formally address the estimation of $\mu_{kt}$ for $1\leq k \leq 4$.

We can now present the two alternative modeling frameworks under which \VaR\ and \ES\ forecasts are generated, denoted as, for ease of reference,  the \emph{additive} and the \emph{multiplicative} models, respectively. We simplify the notation by defining $\val{t}\equiv Q_{\alpha,t}$ and $\es{t}\equiv ES_{\alpha,t}$. Thus, the \emph{additive modeling framework} can be represented as a regression model for the 1-step ahead expected $\alpha$-level of \VaR, $\val{t}$:
\begin{align*}
	r_t&= \val{t}+\eta_t, 
\end{align*}
where the error term $\eta_t$ is controlling the left tail of the conditional distribution of returns, so that, under correct specification of $\val{t}$, the error term $\eta_t$ is such that $F_{\eta}^{-1}(\alpha|\mathcal{I}_{t-1}) = 0$. 

This is a general framework since several models can be derived as special cases by varying the dynamic specifications for $\val{t}$. Noting that a \, $\widehat \cdot$ \, is used to indicate an estimate, $\bar r = T^{-1}\sum_{t = 1}^T r_t$ and $\widehat\Sk^{+}_t$ and $\widehat\Sk^{-}_t$, represent negative and positive skewness:
\begin{equation*}
    \widehat\Sk_{1t}^+ = \widehat\Sk_{1t} \cdot I\{\widehat\Sk_{1t} > 0\},\qquad\qquad
    \widehat\Sk_{1t}^- = -\widehat\Sk_{1t} \cdot I\{\widehat\Sk_{1t} < 0\},
\end{equation*}
in what follows, we investigate three specifications. 

The first is the simple additive form of the VaR being driven only by the lagged observation of an estimator of the integrated volatility ($\widehat\RV_{t-1}$)\footnote{In the absence of jumps, the integrated variance coincides with the quadratic variation that, in turn, diverges from the conditional variance $\RV_t$ by a zero mean error, thus motivating our notation \citep{ande01_jasa}. A set of alternative choices for $\widehat\RV_{t-1}$ will be presented and discussed in Section \ref{s:revsk}.}. 
\begin{equation}
	\mbox{\texttt{add_sim}:}\ \ \val{t}= d_0 + d_1 \widehat\RV^{1/2}_{t-1} + d_2 \val{t-1},\label{eq:add_sim}
 \end{equation}

To account for the potential misspecification in \texttt{add_sim}, we can resort to the Cornish-Fisher (CF) expansion \citep{HilDav1968}, which approximates the quantiles of an unknown non-Gaussian distribution using the information on sample skewness and kurtosis to adjust the value of the corresponding Gaussian quantiles. Considering as an illustration a random variable $X\sim(0,1)$, the CF approximation for the $\alpha$-quantile of $X$ reads as
\begin{equation*}
    X^{CF}_{\alpha}=z_\alpha+\frac{z^2_\alpha-1}{6}Sk +\frac{z^3_{\alpha}-3 z_{\alpha}}{2}Ku-\frac{2z^3_{\alpha}-5 z_{\alpha}}{36}Sk^2, 
\end{equation*}
where $Sk$ and $Ku$ are the usual moment-based sample skewness and kurtosis coefficients of $X$ respectively, and $z_\alpha=\Phi^{-1}(\alpha)$ is the $\alpha$-quantile of a $N(0,1)$ random variable.

Therefore, the second model adds realized negative and positive skewnesses ($\Sk^-_{t-1}$ and $\Sk^+_{t-1}$) and kurtosis ($\Ku_{t-1}$) to the \texttt{add_sim}\footnote{In our approach, we focus on the conditional distribution of returns rather than on their unconditional distribution, as would happen when using the standard CF expansion. Hence, the sample skewness and kurtosis coefficients are replaced by their realized counterparts that provide point estimates of daily conditional skewness and kurtosis.}:
\begin{equation}
	\mbox{\texttt{add_skk}:}\ \ \val{t} = d_0 + d_1 \widehat\RV^{1/2}_{t-1} + d_2 \val{t-1}+(a_1 \widehat\Sk^-_{t-1}+a_2 \widehat\Sk^+_{t-1}+a_3 \widehat\Ku_{t-1}).\label{eq:add_skk}
 \end{equation}
The inclusion of the skewness and kurtosis terms are thus motivated by a data-driven CF expansion, whose coefficients, as it will be later illustrated, can be estimated in a semi-parametric fashion by minimizing a strictly consistent loss function. 

The third model  further extends the \texttt{add_skk} with an asymmetric impact of the integrated volatility in correspondence with returns smaller than their average (leverage effect):
\begin{equation}
	\mbox{\texttt{add_lev}:}\ \ \val{t} = d_0 + d_1 \widehat\RV^{1/2}_{t-1} + d_2 \val{t-1}+(a_1 \widehat\Sk^-_{t-1}+a_2 \widehat\Sk^+_{t-1}+a_3 \widehat\Ku_{t-1}) + d_3 \widehat\RV^{1/2}_{t-1}I\{r_{t-1}\leq \bar r\}.\label{eq:add_lev}
\end{equation} 
By contrast, a \emph{multiplicative modeling framework} can be represented in terms of the following nonlinear regression model 
\begin{align*}
	r_t&= \val{t}\,\eta_t,
\end{align*}
where, under correct specification of $\val{t}$, $\eta_t$ is such that $F_\eta^{-1}(\alpha|\mathcal{I}_{t-1})=1$.

This framework can also be motivated by a simple location-scale representation of the returns process 
\begin{equation*}
r_t=h_t z_t, \qquad\qquad z_t \overset{iid}{\sim} (0,1),
\end{equation*}
where the dynamics of $h_t^2=\Var(r_t|\mathcal{I}_{t-1})$ can be modelled by means of GARCH type models. Under the iid assumption for $z_t$ the 1-step ahead $\alpha$-level \VaR\ of $r_t$ is given by $\val{t} = h_t z_{\alpha}$ where $z_{\alpha}=F_z^{-1}(\alpha)$. Thus, the first multiplicative model is  for the  $\alpha$-level 1-step ahead \VaR is 
\begin{equation}
    \mbox{\texttt{mlt_sim}:}\  \val{t}=h_{t}, \quad
    h^2_{t}=d_0+d_1\widehat\RV_{t-1}+d_2h^2_{t|t-1}.\label{eq:mlt_sim}
\end{equation}
When we allow for a time-varying conditional skewness and kurtosis in the returns distribution, this assumption must be generalized to read
\begin{equation*}
    \val{t}=h_t z_{\alpha,t},
\end{equation*}
where the time variation in the conditional error quantile $z_{\alpha,t}$ is driven by the time-varying conditional skewness and kurtosis values as in the \texttt{mlt_lev} and \texttt{mlt_skk} specifications introduced below. Thus, within the multiplicative modeling framework, we consider the following alternative specifications for the  $\alpha$-level 1-step ahead \VaR
\begin{align}
    \mbox{\texttt{mlt_skk}:}\ \ & \val{t} = h_{t}(a_1 \Sk^-_{t-1}+a_2 \Sk^+_{t-1}+a_3 \Ku_{t-1}), & &
    h^2_{t} = d_0+d_1\widehat\RV_{t-1}+d_2h^2_{t|t-1},\label{eq:mlt_skk}\\
    \mbox{\texttt{mlt_lev}:}\ \  &\val{t} = h_{t}(a_1 \Sk^-_{t-1}+a_2 \Sk^+_{t-1}+a_3 \Ku_{t-1}), & & h^2_{t}=d_0+d_1\widehat\RV_{t-1}+d_2h^2_{t|t-1}+d_3 \widehat\RV_{t-1}I\{r_{t-1}\leq \bar r\}\label{eq:mlt_lev}.
\end{align}
Multiplicative models closely mirror additive models (\ref{eq:add_sim}), (\ref{eq:add_skk}) and (\ref{eq:add_lev}). The \texttt{mlt_sim} is similar to (\ref{eq:add_sim}) and is the simplest specification with only the integrated volatility driving the dynamics of the scale. Further \texttt{mlt_skk} assumes similar to (\ref{eq:add_skk}) in the additional information incorporated in realized skewness and kurtosis that drives the dynamics of the scale. The most complex model \texttt{mlt_lev} also controls for the leverage in $h_t$ in the same fashion as in the model (\ref{eq:add_lev}). 

For both additive and multiplicative frameworks, the \ES\ can be modeled according to two different alternative specifications:
\begin{eqnarray}
    \mbox{\texttt{ES_sim}:}\quad\es{t}&=&\{1+\exp(b_0)\}\val{t}\label{eq:ES_sim},\\
    \mbox{\texttt{ES_skk}:}\quad\es{t}&=&\{1+\exp{\left(b_0 + b_1 \Sk_{t-1}+b_2 \Ku_{t-1}\right)}\}\val{t}.\label{eq:ES_skk}
\end{eqnarray}
Here \texttt{ES_sim} is the simple specification assuming that the \ES\ is a rescaling of \VaR. \cite{tayl2019} shows that this simple specification provides competitive \VaR\ forecasts. More recently, \cite{wang_gerl_chen_2023} have extended the framework proposed in \cite{tayl2019} to allow for separate \VaR\ and \ES\ dynamics as well as for the incorporation of realized measures.

The more complex \texttt{ES_skk} brings the dynamics of the \ES\ to be also driven by the skewness and kurtosis, possibly accounting for the misspecification of \texttt{ES_sim}. Differently from \VaR, in this case, we did not split the skewness into negative and positive. By construction, both \texttt{ES_skk} and \texttt{ES_sim} specifications avoid the crossing of \VaR\ and \ES\ forecasts.

In the additive case, neglecting to model the \ES\ dynamics leads to a pure \VaR\ model. This case, labeled as \texttt{ES_no}, corresponds to a model specification that is close in spirit to a \cite{caviar} CAViaR type one where some realized estimator of the integrated variance replaces the volatility measure based on lagged daily returns.

\subsection{Estimation}
\label{s:esti}
Estimation of the vector $\btheta$  of unknown parameters describing the models for $v_t$ and $e_t$  both in the additive and multiplicative models (\ref{eq:add_sim})-(\ref{eq:ES_skk}) is done semi-parametrically by minimizing a strictly consistent scoring rule,  
\begin{equation}
    \hbtheta=\underset{\btheta}{\arg\min} 
    \sum_{t=1}^{T}S^{(\alpha)}_t,\label{e:optim}
\end{equation}
where $S^{(\alpha)}_t$ is a member of the general class presented by \cite{Fissler2016}, i.e.
\begin{align*}
    S^{(\alpha)}_t \equiv S(\val{t},\es{t}|r_t;\alpha)&=\{I(r_t\leq \val{t})-\alpha\}G_1(\val{t})-I(r_t\leq\val{t})G_1(r_t)+G_2(\es{t})\\
    &\left\{\es{t}-\val{t}+I(r_t\leq\val{t})\frac{\val{t}-r_t}{\alpha}\right\}
    -\zeta_2(\es{t})+a(r_t).\label{e:scorule}
\end{align*}
In the definition of $S^{(\alpha)}_t$, the functions $G_1$, $\zeta_2$, and $G_2$ satisfy the following conditions: $G_1$ is increasing, $\zeta_2$ is increasing and convex, and $G_2=\zeta^'_2$. In particular, setting $G_1(\cdot)=0$, $G_2(x)=-1/x$, $\zeta_2(x)=-\log(-x)$, $a(r_t)=1-\log(1-\alpha)$, leads to the following scoring rule \citep{tayl2019}
\begin{align*}
    AL^{(\alpha)}_t&=\frac{I(r_t\leq\val{t})r_t+\val{t}\{\alpha-I(r_t\leq\val{t})\}}{\alpha \es{t}}+\log(-\es{t})-\log(1-\alpha)\\
    &=\frac{I(r_t\leq\val{t})r_t+\val{t}\{\alpha-I(r_t\leq\val{t})\}}{\alpha \es{t}}-\log\left(\frac{1-\alpha}{\es{t}}\right).
\end{align*}
Adding and subtracting $\alpha r_t$ to the numerator of the second term on the right-hand side of the previous equation, we get
\begin{equation*}
    AL^{(\alpha)}_t=-\log\left(\frac{\alpha-1}{\es{t}}\right)-\frac{(r_t-\val{t})\{\alpha-I(r_t\leq\val{t})\}}{\alpha \es{t} } + \frac{r_t}{\es{t}}.
\end{equation*}
In the simplified expression of $AL^{(\alpha)}_t$ obtained by \cite{tayl2019}, the last term on the right-hand side is dropped. This simplification arises from the assumption that the conditional mean of returns is zero, as shown in their equation (19). Notably, it can be demonstrated that the negative value of $AL^{(\alpha)}_t$ quantifies the contribution of the $t$-observation to a quasi-likelihood function, which is constructed based on the Asymmetric Laplace distribution \citep{tayl2019}.

With the same choices for $G_1$, $G_2$, and $\zeta_2$ as above, but setting $a(r_t)=0$, leads to the zero-degree homogeneous loss used by \cite{pattonetal2019}
\begin{equation*}
    FZ0^{(\alpha)}_t=\frac{I(r_t\leq\val{t})r_t+\val{t}\{\alpha-I(r_t\leq\val{t})\}}{\alpha \es{t}}+\log(-\es{t})-1=
-\frac{I(r_t\leq\val{t})(\val{t}-r_t)}{\alpha \es{t}}+\frac{\val{t}}{\es{t}}+  \log(-\es{t})-1.
\end{equation*}
Therefore, in the case of the model \texttt{EM_no}, where only the \VaR\ is estimated, the objective function is given by the quantile loss:
\begin{equation}\label{e:emloss}
    EM^{(\alpha)}_t = \{\alpha - I(r_t < \val{t})\} \cdot (r_t - \val{t}).
\end{equation}
In what follows, models estimated relying on loss functions $AL^{(\alpha)}_t$, $FZ0^{(\alpha)}_t$ and $EM^{(\alpha)}_t$ are labeled as \texttt{Loss=ALS}, \texttt{Loss=FZ0} and \texttt{Loss=EM}, respectively. Standard errors are computed using the asymptotic theory developed by \cite{caviar}, for pure \VaR\ models, and \cite{pattonetal2019}, for joint \VaR-\ES\ models. Technical details are provided in Section \ref{s:appA}.
 
It is worth noting that optimization in the (\ref{e:optim}) is a challenging task irrespective of what function $S_t^{(\alpha)}$ is chosen, be it either $AL^{(\alpha)}_t$, $FZ0^{(\alpha)}_t$ or $EM^{(\alpha)}_t$. In particular, the optimization of these loss functions is typically strongly dependent upon the chosen initial values. For this reason, we implemented an optimization technique similar to \cite{caviar} which, for ease of reference, we call \emph{complete estimation}. Namely, for each model, we evaluated the objective function on $\mathfrak{n} = 5 \cdot 10^4$ uniformly sampled possible parameter constellations, and among them, we selected the $\mathfrak{m} = 10$ parameter vectors that lead to the smallest objective function values. Selecting each of these $\mathfrak{m}$ vectors as a starting point, we re-estimated the model $\mathfrak{m}$-times iterating between a Nelder-Mead and a BFGS optimizer until convergence is achieved, and the final estimates are those delivering the smallest value of the objective function. In a rolling window forecasting exercise, one may be advised to follow a parsimonious estimation strategy, by using the most recent estimates as the starting point for the next estimation round, at regular intervals.

\section{The underlying process and the derived realized measures}
\label{s:revsk}

Having developed a setup where the theoretical estimators of conditional moments such as $\RV$, $\Sk$, and $\Ku$  are considered within suitable models, we are left with the delicate phase to choose which operational counterparts we can count on at daily frequencies, employing rolling windows and sample statistics. The standard framework starts from a true underlying continuous log-price following a diffusion process, disregarding, for example, the presence of structural breaks: 
\begin{equation*}
    \mathrm{d} X_\mft = \mu(X_\mft)\mathrm{d}\mft +\sigma(X_\mft)\mathrm{d}W_\mft,
\end{equation*}
where, $W_\mft$ represents the standard Brownian motion, $\mu(X_\mft)$ is the drift càdlàg finite variation process, and $\sigma(X_\mft)$ is the time-varying càdlàg volatility function. It is important to note that $\sigma(X_\mft)$ may depend on a separate Brownian motion, which could potentially be correlated with $W_\mft$. This general family encompasses well-known processes such as the Heston or Bates processes (see \cite{heston1993closed, bates1996jumps}). In this context, the parameter $\mft$ represents the continuous temporal component that spans within and across days. 

The second moment $\mu_{2t}$ is known as the \emph{integrated variance}, an object of paramount importance to researchers and practitioners. By utilizing the aforementioned process over a one-day interval $[t-1d, t]$, the integrated variance can be computed as $\int_{t-1d}^t \sigma^2(u)\mathrm{d}u$. 

The temporal component then needs to be somehow aggregated to get the daily estimators for the relevant moments: in this respect, we ground ourselves in the massive literature that considers the market activity of a day (using the same index $t \in \{1, \ldots, T\}$) between opening and closing to be divided into regularly spaced intervals $i \in \{0, \ldots, N\}$. We then take the high-frequency log-prices $x_{t,i}$ as the elementary information, to be converted into $r_{t, i} = x_{t,i} - x_{t, i-1}$, the corresponding \emph{intraday} log-returns, $i=1,\ldots,N$.

The overwhelming attention of the literature was devoted to the design of consistent estimators of the integrated variance $\mu_{2t}$ of the continuous process over a discrete interval \citep{andersen2010parametric}, with specific care devoted to departures from the standard framework (e.g. jumps) or to the nature of observed prices which are affected by trading mechanisms (so-called market microstructure). There exists a range of options for researchers and practitioners alike seeking to estimate these quantities accurately and efficiently. Starting from the realized variance \citep{AndersenBoll98},
\begin{equation*}
    \widehat\mu_{2t}^{RV} = \sum_{i = 1}^{N} r_{t,i}^2,
\end{equation*}
other widely used estimators of the integrated volatility are the, proposed in \cite{BarndorffNielsen_Bipower_2004} and \cite{andersen2012jump}, bipower variation $\widehat\mu_{2t}^{BPV} = \frac{\pi}{2}\frac{N}{N-1}\sum_{i=1}^{N-1}|r_{t,i}||r_{t,i+1}|$, or the upside and downside semivariances $\widehat\mu_{2t}^{SVPOS}= \sum_{i=1}^{N} r_{t,i}^2 \ \cdot \ I \{ r_{t,i} > 0\}$ and $\widehat\mu_{2t}^{SVNEG} = \sum_{i=1}^{N} r_{t,i}^2 \ \cdot \ I \{ r_{t,i} < 0\}$ developed in \cite{Barndorff_Semivariance_2008} and \cite{Bollerslev_SemiCovar_2020}. 

 Barring a horse race among the many estimators of $\mu_{2t}$ available, we limit ourselves to a single choice, and our preference goes to the median estimator 
\begin{equation*}
    \widehat\mu_{2t}^{MED}=\frac{\pi}{6-4\sqrt{3}+\pi}\frac{N}{N-2}\sum_{i=2}^{N-1} \mbox{med}(|r_{t,i-1}|,|r_{t,i}|, |r_{t,i+1}|)^2,
\end{equation*}
proposed by \cite{andersen2012jump}, because of its documented robustness properties.

Several new estimators for \emph{higher-order} moments have emerged in recent years. While these estimators do not directly estimate daily skewness or kurtosis, they instead estimate the integrated third or fourth power of intraday returns or the averaged jump component. Empirical evidence suggests that these estimators can be informative in predicting cross-sectional next week's stock returns or in forecasting RV at medium- to long-term horizons, as demonstrated by \cite{MeiLiuMaChen2017}. The simplest estimator of the integrated $k$-order moments was proposed by \cite{amaya2015}, shadowing the relationship between the realized variance and the integrated volatility (case $k = 2$).
\begin{equation*}
	\widehat{\mu}_{kt}^{ACJV} = \sum_{i = 1}^{N} r_{t,i}^k.
\end{equation*}
\cite{amaya2015} demonstrate the estimator's consistency, which asymptotically captures only the jump component and the average jump size but does not capture skewness arising from the leverage effect and heavily depends on the sampling frequency. Later \cite{LiuWangLiu2014} derived asymptotic properties of the \cite{amaya2015} estimator and developed their own measures of realized skewness accounting for market microstructure noise. Another extension was provided by \cite{ChoeLee2014}, who showed that the daily third moment is proportional to the quadratic covariation between the squared return and the return process, and the fourth moment is proportional to the quadratic variation of the squared return process with some additional cross‐terms. 

Based on some preliminary analysis, our choice for the realized skewness and kurtosis falls on the estimators by 
\cite{neuberger2013} and \cite{neuberger2020}:
\begin{align*}
	\widehat{\mu}_{3t}^{NP} & = \frac{1}{\tau} \sum_{j = 0}^{\tau-1} \sum_{i = 1}^{N} \left(r_{t-j,i}^3 + 3y_{t-j,i-1}^* r_{t,i}^2\right),
	&\widehat{\mu}_{4t}^{NP} & = \frac{1}{\tau} \sum_{j = 0}^{\tau-1} \sum_{i = 1}^{N} \left(r_{t-j,i}^4 + 4 y^*_{t-j,i-1} r_{t-j,i}^3 + 6z_{t-j,i-1}^*r_{t-j,i}^2\right),
\end{align*}
where $y_{t,i-1}^* = \frac{1}{N}\sum_{j=1}^{N}(x_{t,i-1} - x_{t,i-j})$ and $z^*_{t,i-1} =  \frac{1}{N}\sum_{j=1}^{N}(x_{t,i-1} - x_{t,i-j})^2$ measure local (daily) trends in simple and squared log-prices. Similar to \cite{ChoeLee2014}, they assume that the conditional mean of the returns is zero.  Also, in what follows, we choose $\tau = 5$.  

We note that the estimators for realized skewness and kurtosis sometimes produce outliers that can significantly affect the performance of \VaR\ and \ES\ models. To address this issue, we applied a filter that removes estimated skewness and kurtosis values falling outside the ranges of $(-15;15)$ and $(0;20)$, respectively. Outliers excluded from our analysis are then smoothed out using interpolation techniques accounting for autocorrelation.

%
%

\section{Empirical evidence}
\label{s:appli}

\subsection{Data and forecasting design}
In this section, we present the results of our setup to a very large panel of 960 U.S. stocks traded on the New York Stock Exchange (NYSE), included in the S\&P500 index at various times over the considered period. The list of stocks can be found in Web Appendix \textit{List of Tickers}. The original dataset for each stock consists of intra-daily prices adjusted for stock splits and dividends sampled every 5 minutes. We focus only on regular trading hours, from 9:30 am to 4:00 pm, resulting in 78 observations for each trading day. The stocks have different timespans, starting within a range between 1998-01-02 and 2016-10-11, and ending between 1998-01-09 and 2017-02-09. Furthermore, to ensure an adequate sample size, we limit our analysis to assets with a continuous record of at least 500 daily observations. This reduces the cross-sectional size of our sample to 823 assets (marked in the Web Appendix \textit{List of Tickers} in italics).

Our empirical strategy consists of two main steps. In the first, we conduct a full-sample analysis to assess the performance of various models in fitting \VaR\ and \ES. In the second step, we focus on the out-of-sample forecasting performance using a rolling window approach. We consider three estimation windows: 500, 1000, and 2000 days. For the out-of-sample analysis, we include assets with a continuous record of daily pricing observations from the start date of our sample to its end, 2017-02-09. In this case, we were able to obtain one-step ahead predictions for the dates 2000-01-04 -- 2017-02-09, for $w = 500$, 2002-01-03 -- 2017-02-09 for $w = 1000$, and 2005-12-21 -- 2017-02-09 for $w = 2000$. 406 of the original 960 stocks meet this criterion and are included in the out-of-sample analysis (marked in boldface in the Web Appendix \textit{List of Tickers}).

The model universe considered for both the full-sample and out-of-sample analysis includes all the specifications presented in Section \ref{s:mods}. As discussed, each of these is coupled with three different \ES\ specifications, \texttt{ES_sim}, \texttt{ES_sk}, and \texttt{ES_no}, for a total of 18 different models. Implementing the procedures discussed in Section \ref{s:esti}, each model is estimated for three different risk levels, $\alpha \in \{0.01,0.025,0.05\}$.

\subsection{Analysis of the in- and out-of-sample losses and coverage}
Before delving into the assessment of the model performances through the various tests conducted on the extensive dataset, it is first useful to visually assess their in- and out-of-sample coverage.
Figure \ref{fig:insampleCov} presents a comprehensive overview of the aggregated information across all datasets, three coverage levels, and all models for in-sample performance. Each model is estimated for every dataset, and the in-sample empirical coverage ($\hat\alpha$) is calculated. The models are represented by row blocks in the figure, with corresponding names on the y-axis, such as ``\texttt{VaR=m_lev, ES=no, Loss=EM}''. Within each block, three box plots display the coverage for all datasets, with blue indicating $\alpha = 0.01$, green representing $\alpha = 0.025$, and red representing $\alpha = 0.05$. These levels are also depicted by vertical dashed lines. All the models exhibit similar behavior and, on average, achieve the desired coverage level, albeit with slight variations. Simpler models generally exhibit less variability. Some cases encountered convergence difficulties, leading to the inability to estimate certain models. The right panel of Figure \ref{fig:insampleCov} shows the fraction of such problematic cases, consistently below 3\%.

A similar analysis has been conducted for out-of-sample coverage, utilizing three different window sizes of 500, 1000, and 2000 days. Aggregated results are presented in Figure \ref{fig:outofsampleCov}. In addition to the three colors representing coverage levels ($\alpha = 0.01$, $\alpha = 0.025$, and $\alpha = 0.05$), varying color intensities indicate window size (lightest shade = 2000 days; darkest = 500 days).\footnote{Fewer models are considered in the out-of-sample analysis, excluding "\texttt{ES=SkKu}" due to computational complexity and "\texttt{Loss=FZ0}" due to its similar behavior to "\texttt{Loss=ALS}".} The out-of-sample results reveal a less favorable situation than the in-sample, as all models tend to overestimate the coverage on average, less severely so with a wider rolling window. Surprisingly, the variance also increases in this scenario. This can be attributed to longer intervals containing more diverse data from potentially different underlying models, thus imperfectly capturing future behavior. Despite these nuances, all models demonstrate similar behavior based on simple visual inspection. Additionally, Figure \ref{fig:outofsampleLosses} provides aggregated loss information. It is evident that both the values and spreads of the loss function decrease with larger sample sizes.

\subsection{Evaluation metrics}
Following the practice by researchers and risk managers, the in- and out-of-sample performances of the dynamic models\footnote{It should be noted that performing a complete estimation for all rolling windows in the out-of-sample exercise has been highly time-consuming. As a result, we perform a full estimation for the initial window and subsequently at 500-day time intervals, and, instead of repeating the complete estimation process for each subsequent window, we update the parameters at regular intervals. To accomplish this, we perform parameter optimization every 50 observations, starting from the results obtained in the previous step. This parameter updating allows us to refine the estimation without repeating the entire process. In all other rolling windows, we maintain the parameters obtained from the previous window.} for \VaR\ and \ES\  presented in Section \ref{s:mods} are firstly assessed using some diagnostic tests, whose technical details are summarized in Appendix \ref{s:hit}-\ref{s:patton} for the reader's convenience.

To assess the in-sample \VaR\ estimation performance, we consider the in-sample Dynamic Quantile (DQ) test by \cite{caviar}. In particular, we consider the test in its conditional coverage and independence 
versions as described in \cite{dumietal2012}. The asymptotic theory for these tests was originally derived by \cite{caviar} for the pure CAViaR models. Therefore, the results presented hereafter refer to the \texttt{ES_no} case only, involving CAViaR models estimated by minimizing the aggregated quantile loss.

While the in-sample DQ test assesses the goodness-of-fit of CAViaR models, its out-of-sample counterpart can be seen as a general test for evaluating the statistical properties of a set of \VaR\ forecasts, regardless of the model. This includes testing for unbiasedness, independent hits, and the independence of quantile estimates, as outlined by \cite{caviar}. In our case, the out-of-sample DQ (OOS-DQ) test can effectively evaluate the properties of the \VaR\ forecasts generated by joint dynamic \VaR-\ES\ models.

Further, we jointly assess the statistical accuracy of \VaR\ and \ES\ estimates, both in and out-of-sample, employing two regression-based testing procedures, i.e., the regression-based calibration tests by \cite{pattonetal2019}, henceforth PZC, and the ESR test by \cite{bay_dim_2020}. The former includes separate calibration tests for \VaR\ and \ES\ while the latter test is specific for \ES\ diagnostics. Moreover, the PZC tests are based on OLS auxiliary regression equations where the standardized generalized residuals \citep[as in][]{pattonetal2019}  are regressed on their past values as well as on \VaR\ and \ES\ forecasts, respectively.

The ESR approach, instead, is based on  three separate test statistics: the Auxiliary, the Strict and the Strict Intercept Backtest, which can be seen as an extension of Mincer-Zarnowitz regression to a semi-parametric setting, relying on the minimization of the consistent loss functions proposed by \cite{Fissler2016}. The Auxiliary and Strict test statistics are computed regressing returns on the \ES\ forecasts and test the \ES\ coefficients for joint $(0, 1)$ values. Specifically, the Auxiliary test requires an auxiliary \VaR\ forecast, while the Strict Intercept tests whether the expected shortfall of the forecast error $(r_t-ES_t)$ is zero.\footnote{\cite{bay_dim_2020} also consider a one-sided version of the Strict Intercept that is particularly useful for regulatory evaluations. Since our main interest is simply in the assessment of forecasting accuracy, in this paper we only consider the two-sided version of the test.}

Finally, the out-of-sample forecasting accuracy of each of the models is assessed by comparing the average values of the FZ0 loss achieved over the forecasting period.

\subsection{Testing the properties of the risk estimates}
In this section, we analyze the properties of the in-sample risk estimates over the full-sample for the set of 823 assets, postponing to a later subsection the generation of out-of-sample risk forecasts. Also, for the sake of brevity, we only discuss results for $\alpha=0.01$, while the results for $\alpha=0.025$ and $\alpha=0.05$ are contained in the Web Appendix \textit{Tables W.1 and W.2}.

The \emph{In-sample DQ} section of Table \ref{tab:1} reports the results of the in-sample DQ test in its conditional coverage ($CC-DQ_{IS}$) and independence ($ID-DQ_{IS}$) versions, respectively. For each model, the table provides the non-rejection frequency at the $5\%$ significance level. Higher values in the table indicate better performance, as they correspond to models less frequently rejected by the tests.

Although the $CC-DQ_{IS}$ provides a comprehensive evaluation of the risk estimation performance, it has a \textit{portmanteau} nature, which overlooks the clustering features of the hit series and neglects their coverage properties. By contrast, the $ID-DQ_{IS}$ test offers a complementary perspective to the previous test, since it focuses explicitly on clustering. Combining the information from both tests makes it possible to gain deeper insight into the reasons behind any model underperformance.
The test findings for $\alpha=0.01$ (similar results hold for the other risk levels) can be summarized as follows:
\begin{itemize}
    \item[] $CC-DQ_{IS}$ : multiplicative models outperform their additive counterparts with the \texttt{mlt_lev} resulting the best model at all risk levels. This model is not rejected at the 5\% level in approximately 70\% of cases, closely followed by the \texttt{mlt_skk}. The \texttt{mlt_sim} yields slightly lower rates than models incorporating information on realized skewness and kurtosis. The non-rejection frequency of additive models is much lower, being on average close to $20\%$, with the highest rate being recorded for the \texttt{add_sim} model. 
    \item[] $ID-DQ_{IS}$: multiplicative and additive models are characterized by similar performances, suggesting that the high rejection rate of the latter class is mostly due to lack of coverage rather than to hit clustering.
\end{itemize}

Second, to appreciate the contribution of the additional information in the form of realized skewness and kurtosis in the \texttt{skk} models, in the \emph{Wald test} section of Table \ref{tab:1} we assess the significance of the skewness and kurtosis coefficients involved in the \VaR\ and \ES\ dynamics, respectively. Specifically, we test the null $a_1 = a_2 = a_3=0$, for \VaR, and $b_1 = b_2 = 0$, for \ES,  against a two-sided alternative. 

Again, the test results in terms of empirical non-rejection frequencies are summarized over the panel of assets considered. For the plain CAViaR models, the null $a_1 = a_2 = a_3 =0$ is almost always rejected at the usual 5\% significance level, for all risk levels considered. Differently, for joint  \VaR-\ES\  models, the non-rejection frequency increases with the risk level $\alpha$. Namely, the percentage of non-rejections is close to 0 for $\alpha = 0.01$ but it increases to values  up to $\approx 40\%$ for $\alpha = 0.05$. 
The discrepancy between non-rejection frequencies for pure \VaR\ and joint \VaR-\ES\ models is likely to be due to the fact that, for each class of models, testing is based on a different asymptotic distribution: we rely on the theory derived by \cite{caviar}, for the EM loss, and on \cite{pattonetal2019}, for ALS and FZ0. The test results are only marginally affected by the choice of the joint loss, AL or FZ0, used for estimation.

Moving to the analysis of \ES\ dynamics, we find that the non-rejection frequencies of the null $b_1 = b_2 = 0$ are substantially higher than the values observed for \VaR\ parameters and are clearly affected by the risk level. Namely, they approximately lie in the range 49\%-61\%, for $\alpha=0.01$, 55\%-72\%, for $\alpha=0.025$, and 62\%-79\%, for $\alpha=0.05$. Results are very close for models based on ALS and FZ0 losses. Overall, we conclude that the inclusion of realized skewness and kurtosis measures in the \ES\ equation is less strongly supported than for the \VaR.

Next, we focus on the in-sample PZC and ESR calibration tests. First, the \emph{Calibration test} section  of Table \ref{tab:1} for $\alpha = 0.01$ (see Web Appendix \textit{Tables W.1 and W.2} for $\alpha= 0.025$ and $\alpha= 0.05$) reports the results of the former tests for \VaR\ and \ES. The main findings arising from the table can be summarized as follows:
\begin{itemize}
    \item for $\alpha \geq 0.025$ (Tables W.1 and W.2), all models yield remarkably good non-rejection frequencies, with values ranging from 72\% to 94\%.
    \item For both \VaR\ and \ES, we record a decay of the non-rejection frequency at the 0.01 risk level (Table \ref{tab:1}). This is particularly relevant for \VaR\ since $\alpha=0.01$ is the mandatory level indicated by the Basel Committee.
    \item Models based on ALS and FZ0 losses return very close performances.
    \item Comparing simpler models (\texttt{*_sim}) with more complicated specifications (\texttt{*_skk} and \texttt{*_lev}), we find that there is no clear winner but the ranking depends on the functional form and risk level.
\end{itemize}

Finally, to assess the "calibration" of \ES\ forecasts, the \emph{ES calibration test} section in Table \ref{tab:1} for $\alpha = 0.01$ (see Web Appendix \textit{Tables W.1 and W.2} for $\alpha= 0.025$ and $\alpha= 0.05$) reports the non-rejection frequencies of the three ESR tests proposed by \cite{bay_dim_2020}\footnote{The tests were implemented using the \texttt{esback} \texttt{R}-library provided by the same authors, freely available from CRAN at the URL: https://cran.r-project.org/web/packages/esback/index.html}. It is worth noting that due to numerical problems in the computation of the test statistic, this could not be computed for some of the assets in our panel, in addition to those that had been previously excluded due to convergence issues in the estimation of the reference risk models: the number of valid assets for each configuration, determined by a combination of available models and risk levels, ranges from a minimum of 644 to a maximum of 796 assets out of 823.  

Compared to the calibration test by \cite{pattonetal2019}, the ESR reveals a much lower discriminatory power returning non-rejection frequencies very close to unity for all models and risk levels. Again, we do not report any apparent differences in model performances based on the ALS and FZ0 losses. 

\subsection{Out-of-sample forecasting comparison}
\label{sec:oos}
This section presents the results of the out-of-sample forecasting analysis. First, the performance of the models under analysis is assessed by computing the following test statistics and diagnostics over the out-of-sample period
\begin{itemize}
    \item  DQ tests for independence and conditional coverage
    \item \VaR\ and \ES\ calibration tests by \cite{pattonetal2019} 
    \item ESR tests for \ES\ calibration by \cite{bay_dim_2020}.
\end{itemize}
As in the previous section, test results across the whole panel of assets are summarized in terms of empirical non-rejection frequencies. Also, we only discuss results for $\alpha=0.01$ in Table \ref{tab:2} while results for $\alpha=0.025$ and $\alpha=0.05$ have been reported in the Web Appendix \textit{Tables W.3 and W.4}.

Similarly to what was observed in the full sample analysis, in a limited number of cases it has not been possible to calculate the $p$-values of ESR tests due to failures in the estimation of the auxiliary regression model underlying the test. Overall, depending on risk level, specific test of interest, and sample size, the available number of stocks has been found to range between 371 and 406 out of 406 potentially available stocks. 

The findings of the analysis can be succinctly summarized as:
\begin{itemize}
    \item DQ tests: the non-rejection frequencies are very low for the shortest estimation window $T=500$ but they tend to increase with the sample size although, even for $T=2000$, they barely exceed $40\%$, for independence tests, only in a few isolated cases. Overall some stylized facts arise. Plain \VaR\ models on average perform better than joint \VaR-\ES\ models while the inclusion of information on skewness and kurtosis does not bring any evident advantages.
    \item \VaR\ calibration tests: the performances are very poor for the shortest sample size $T=500$ but tend to improve as $T$ increases. The model performances also depend on the value of the risk level $\alpha$ with the best results obtained for $\alpha=0.025$. In terms of model specifications, \texttt{add_sim} and \texttt{mlt_sim} yield the highest non-rejection frequencies that exceed $70\%$  for T=2000 and $\alpha=0.025$ when the EM loss is used. When comparing plain \VaR\ and joint \VaR-\ES\ models, there are no clear performance gaps.
    \item \ES\ calibration tests: the results are qualitatively not different from what was observed for the \VaR\ tests. Hence, similar considerations hold.
    \item ESR tests: the performance of the ``strict'' and ``auxiliary'' tests improves as the sample size increases although the performance gap across different sample sizes is less evident than for the other regression-based \VaR\ and \ES\ calibration tests. As above, even in this case, we record the best performances for the \texttt{add_sim} and \texttt{mlt_sim} models reaching, in some cases, non-rejection frequencies close to $80\%$. As far as the ``strict intercept test'' is concerned, the differences across different models and sample sizes are much less evident and the non-rejection frequency is $>80\%$ in all instances. Again, the information on realized skewness and kurtosis does not appear to lead to improvements in terms of forecasting performances.
\end{itemize}

Finally, we assess and compare the forecasting accuracy of the different models based on the out-of-sample values of the following strictly consistent scoring functions: quantile loss (E) for \VaR\ and AL-score for joint \VaR\ and \ES\ forecasting (ALS).
In terms of median loss (Table \ref{tab:medianloss}), the multiplicative model without skewness and kurtosis information (\texttt{mlt\_sim}) achieves the minimum loss value in most cases for both quantile and ALS scoring functions. It is only slightly outperformed by its additive counterpart (\texttt{add\_sim}) in one instance for pure \VaR\ models and in two instances for joint forecasts of \VaR\ and \ES. A similar trend is observed when considering average ranks (Table \ref{tab:medianranks}). The \texttt{mlt\_sim} model consistently delivers the minimum average rank, except in the case of joint \VaR\ and \ES\ forecasts at the 0.05 level and for $T=1000$, where it ranks second, closely following the \texttt{add\_sim} model that also does not use skewness and kurtosis information.

In conclusion, the key insights from the assessment of forecasting performance can be summarized as follows:
\begin{itemize}
\item Incorporating information on realized skewness and kurtosis does not enhance forecasting accuracy;
\item Simpler models are preferable to more complex ones, as the latter are more vulnerable to computational issues;
\item The \emph{multiplicative} specification is generally preferable to the more popular \emph{additive} approach.
\end{itemize}

\section{Concluding remarks}
\label{s:conc}
In this paper, we have presented a forecasting comparison of several semi-parametric risk forecasting models. Our work presents some important elements of novelty and potential interest for practitioners and researchers alike. First, the comparison is based on an unusually large set of 823 stocks: to the best of our knowledge, there are no other contributions relying on such a large dataset in the tail-risk forecasting literature. Also, the availability of such a rich data environment has a positive impact on the reliability of the regularities that emerge from the empirical analysis, giving them a good degree of external validity.
Second, we assess the potential contribution coming from considering information on some recently proposed realized skewness and kurtosis measures. Third, we provide deeper insight into the selection of the functional form of the semi-parametric model used to generate forecasts. 

The results of our analysis clearly indicate that, at the forecasting stage, simple models should be preferred to more complicated ones with a preference for multiplicative GARCH-type specifications. Realized skewness and kurtosis measures do not apparently provide valuable information for improving the accuracy of tail risk forecasts even if in most cases, their coefficients turn out to be significant in the full-sample analysis. By the same token, they may prove useful in generating improved density forecasts, a task that we leave for future research.

When we shift the focus to the functional form of the dynamic risk model, an interesting and original finding from our extensive empirical investigation is that the standard \emph{CaViaR-like} additive model specification outperformed by the less commonly used (in a semi-parametric framework) \emph{GARCH-like} multiplicative parameterization.

\bigskip
\bibliographystyle{apalike}
\bibliography{literature}

\newpage

\section*{Appendix A: Asymptotic distribution of the estimators}
\setcounter{figure}{0} \renewcommand{\thefigure}{A.\arabic{figure}}
\setcounter{table}{0} \renewcommand{\thetable}{A.\arabic{table}}
\setcounter{subsection}{0} \renewcommand{\thesubsection}{A.\arabic{subsection}}

\label{s:appA}
\subsection{Standard errors estimation for pure \VaR\ models}
\label{s:A1}
The theoretical results presented in this section are based on \cite{caviar} and assumptions therein. It is worth noting that, although \cite{caviar} focuses on additive CAViaR models, their framework readily applies to the multiplicative \VaR\ models class. In the following, we assume that the return process $r_t$ has conditional $\alpha$-quantile given by $\val{t}(\btheta_0)$. The estimated \VaR, $\val{t}(\estiv)$, is obtained by replacing $\btheta_0$ with the minimizer of the aggregated quantile loss function. Applying the results in Section 4 of \cite{caviar},  $\estiv$ can be shown to be consistent and asymptotically normal. In particular, we have
\begin{equation*}
    \sqrt{T}A^{-1/2}_T D_T (\estiv-\btheta_0)\overset{d}{\to}MVN(\mathbf{0},I)\quad \textrm{as } T\to \infty,
\end{equation*}
where
$A_T= \E\left[T^{-1}\alpha(1-\alpha) \sum_{t=1}^{T}\nabla^' \val{t}(\parv_0)\nabla \val{t}(\parv_0)  \right]$, $D_T= \E\left[T^{-1} \sum_{t=1}^{T} h_t(0|\mathcal{I}_{t-1})\nabla^' v_t(\boldsymbol{\parv_0})\nabla \val{t}(\parv_0)\right]$, $h_t(0|\mathcal{I}_{t-1})$ is the conditional density of $\eta_t=r_t-\val{t}$ evaluated at 0 and $\nabla \val{t}(\parv)=\partial{\val{t}(\parv)}/ \partial \parv$. Consistent estimates of $A_T$ and $D_T$ can be then obtained as follows
\begin{equation*}
    \widehat{A}_T=T^{-1}\alpha(1-\alpha) \nabla^\top v(\parv_0) \nabla (v\parv_0),
\end{equation*}
where
$\nabla v(\boldsymbol{\theta_0})$ is the ($T\times p$) matrix whose $t$-th row is $\nabla^\top \val{t}(\parv_0)$, and
\begin{equation*}
    \widehat{D}_T=(2 \, T \, \widehat{c}_T)^{-1}
    \sum_{t=1}^{T}I(|r_t -\val{t}(\estiv)|<\widehat{c}_T)
    \nabla' \val{t}(\estiv)
    \nabla \val{t}(\estiv).
\end{equation*}
Analytical expressions for the elements of $\nabla \val{t}(\estiv)$ have been derived and reported in Web Appendix \textit{Derivatives}. 
Following \cite{caviar}, the bandwidth $\widehat{c}_T$ is set as: $\widehat{c}_T=40$, for $\alpha=0.01$, $\widehat{c}_T=60$, for $\alpha=0.05$. For the case $0.01<\alpha<0.05$, which is not considered in the paper by \cite{caviar}, we estimate the bandwidth by linear interpolation. So the final estimated asymptotic variance and covariance matrix of $\hbtheta$ is computed as $\widehat{\Sigma}_{\hbtheta}=\frac{1}{T}\,(\alpha)\,(1-\alpha) \widehat{D}^{-1}_T \, \widehat{A}_T \, \widehat{D}^{-1}_T$ and estimated standard errors are computed as $\widehat{se}(\hbtheta)=\sqrt{\operatorname{diag}\left(\widehat{\Sigma}^{-1}_{\hbtheta}\right)}$. 

Letting $\mathbf{a} \subset \btheta$, the above results can be used to test $\mathbf{a}=\mathbf{0}$. Note that we can write $R \btheta = \mathbf{a}$ where
\begin{equation*}
    R=\left( 
    \begin{array}{c}
    \mathbf{0}_{n-2,n}\\
    \hline
     \begin{array}{c | c}
     \mathbf{0}_{2,n} & I_2
     \end{array}
    \end{array}
    \right)
\end{equation*}
$\mathbf{0}_{m,n}$ indicate a $(m \times n)$ matrix of zeroes and $I_n$ be a ($ n \times n$) identity matrix. Relying on the asymptotic normality of  $\boldsymbol{\widehat{\theta}}$, it can then be shown that, under the null $R \btheta= \mathbf{0}$, $(R \boldsymbol{\widehat{\theta}})^\top(R \widehat{V}_T R^\top)^{-1} (R\boldsymbol{\widehat{\theta}})\underset{d}{\to}\chi^2_2$, as $T \to \infty$.

\subsection*{Standard errors for joint \VaR-\ES\ models}

 \cite{pattonetal2019} prove consistency and asymptotic normality for the estimator $\estij{}$. The theoretical results presented in this section rely on the theory developed in \cite{pattonetal2019} and assumptions therein. In the presentation of the following results, we assume that the return process $r_t$ has theoretical $\alpha$-level \VaR\ and \ES\ given by $\val{t}(\parj_0)$ and $\es{t}(\parj_0)$, respectively. The estimated \VaR\ and \ES, $\val{t}(\estij)$ and $\es{t}(\estij)$, are obtained by replacing $\parj$ with the minimizer of the strictly consistent loss function used for estimation. It is worth noting that, although the results in  \cite{pattonetal2019} are derived for estimators based on the FZ0 loss function, they can be immediately extended to estimators obtained by minimizing different loss functions, such as the ALS.

In particular, the asymptotic distribution of $\estij$ is given by 
\begin{equation*}
    \sqrt{T}A^{-1/2}_0 D_0 (\estij-\parj_0)\overset{d}{\to}MVN(\mathbf{0},I)\quad \textrm{as } T\to \infty.
\end{equation*}
Consistent estimates of $A_0$ and $D_0$ can be obtained as follows
\begin{align}
    \widehat{A}_T&=T^{-1} \sum_{t=1}^{T}\lambda_t(\estij)\lambda^\top_t(\estij)\\
    \widehat{D}_T&=T^{-1} \sum_{t=1}^{T}\left\{\frac{1}{2 c_T}
    I(|r_t - \val{t}|< c_T)
    \frac{\nabla^\top\es{t}(\estij)\nabla \val{t}(\estij)}{-\alpha \es{t}(\estij)}
    +
    \frac{\nabla^\top\es{t}(\estij)\nabla \es{t}(\estij)}{ \es{t}^2(\estij)}
    \right\}
\end{align}
where the bandwidth $c_T$ is set equal to $T^{-1/3}$, as suggested by \cite{pattonetal2019} and $\lambda_t(\parj)=\partial{L^{(\alpha)}_t}/\partial{\parj}$, with $L^{(\alpha)}_t$ denoting the strictly consistent loss function used for estimation.  When $L^{(\alpha)}_t\equiv FZ0^{(\alpha)}_t$, $\lambda_t(\parj)$ is given by
\begin{eqnarray*}
    \lambda_t(\parj) &=&\der{FZ0^{(\alpha)}_t}{\parj}=
    \nabla^\top\val{t}(\parj)\frac{1}{-\es{t}(\parj)}
    \left[\frac{1}{\alpha}I\{\val{t}(\parj)\} -1\right]\\
    &+&\nabla^\top\es{t}(\parj)\frac{1}{\es{t}(\parj)^2}
    \left[\frac{1}{\alpha}I\{\val{t}(\parj)\}\{\val{t}(\parj)-r_t\}-\val{t}(\parj)+\es{t}\right].
    \label{e:dfz0}
\end{eqnarray*}
If the ALS loss is used, the formula above becomes
\begin{eqnarray*}
    \lambda_t(\parj)&=&\der{ALS^{(\alpha)}_t}{\parj}=
    \nabla^\top\val{t}(\parj)\frac{1}{-\es{t}(\parj)}
    \left[\frac{1}{\alpha}I\{\val{t}(\parj)\} -1\right]\\
    &+&
    \nabla^\top\es{t}(\parj)\frac{1}{\es{t}(\parj)^2}
    \left[\frac{1}{\alpha}I\{\val{t}(\parj)\}\{\val{t}(\parj)-r_t\}-\val{t}(\parj)+\es{t}(\parj)+r_t\right]
\end{eqnarray*}
that differs from (\ref{e:dfz0}) for the return $r_t$ in the last term on the RHS. Analytical expressions for the elements of $\nabla \es{t}(\estij)$  have been derived and reported in Web Appendix \textit{Derivatives}.
Finally, an estimate of the asymptotic variance and covariance matrix of $\parj$ is then given by
    $\widehat{\Sigma}_{\hbtheta}=T^{-1}\widehat{D}^{-1}_T \, \widehat{A}_T\, \widehat{D}_T$
and standard errors are computed as $\widehat{se}(\hbtheta)=\sqrt{\operatorname{diag}\left(\widehat{\Sigma}^{-1}_{\hbtheta}\right)}$.

\section*{Appendix B: Diagnostic tests for \VaR\ and \ES}
\setcounter{figure}{0} \renewcommand{\thefigure}{B.\arabic{figure}}
\setcounter{table}{0} \renewcommand{\thetable}{B.\arabic{table}}
\setcounter{subsection}{1} \renewcommand{\thesubsection}{B.\arabic{subsection}}
\label{s:appB}

\subsection{In-sample hit test}
\label{s:hit}

The in-sample hit test is proposed by \cite{caviar} as a model-based diagnostic test for detecting misspecified CAViaR models. The test relies on the \emph{Hit} variables defined as $Hit_t=I(v_t)-\alpha$. Replacing $v_t$ by its estimated counterpart $\hat{v}_t$, the estimated hits are obtained as $\hHit_t=I(\hat{v}_t)-\alpha$ and stacked together into $\hbHit=(\hHit_{q+1},\ldots,\hHit_T)^\top$. Then, letting $\parv$  denote the vector of CAViaR coefficients, define $\bX_t(\parv_0)$ as $\bX_t(\parv_0)=(Hit_{t-1}, \ldots, Hit_{t-q},\bz^\top_{t-1})$, where $\bz_{t}$ is a ($m\times 1$) vector of $\mathcal{I}_{t}$ measurable instruments, for $t = q + 1, \ldots, T$. For example, $\bz^\top_{t-1}$ could include estimated past \VaR\ values or realized measures of skewness and kurtosis. Let then $\bX(\btheta_0)$ be the matrix whose generic row is given by $\bX_t(\parv_0)$, so that $\bX(\parv)$ is of dimension $(T - q)\times(q + m)$. Then, define $(q + m) \times (T - q)$-dimensional matrix
\begin{equation*}
    \bM_T=\bX^\top(\parv_0) - \E\left\{T^{-1}\bX^\top(\parv_0) \bH \nabla v_t(\parv_0)\right\}D^{-1}_T,
\end{equation*}
where $\bH$ is a diagonal matrix with diagonal entries given by $h_t(0|\mathcal{I}_{t-1})$ as defined in \ref{s:A1}. Under the assumptions from \cite{caviar}, we have
\begin{equation*}
    \left\{ \alpha \,(1-\alpha) \E \left(T^{-1} \bM_T \bM^\top_T\right)\right\}^{-1/2} T^{-1/2}\bX^\top(\hparv)\, \hbHit \overset{d}{\to} N(\mathbf{0}_{{q + m}},\mathbf{I}_{{q + m}}),
\end{equation*}
%
where $\mathbf{0}_{q + m}$ is the $q+m$ vector of zeros, and $\mathbf{I}_{q + m}$ is the $q+m$ dimensional identity matrix. In the above result, the matrix $\bM_T$ can be estimated as
\begin{equation*}
    \hbM_T=\bX^\top(\hparv)-\left\{(2 T \widehat{c}_T)^{-1} \sum_{t=1}^{T}I(|r_t-v_t(\hparv)|<\widehat{c}_T)\,\bX^\top_t(\hparv) \nabla v_t(\hparv) \right\}\widehat{D}^{-1}_T \bg^\top(\hparv),
\end{equation*}
with $\bg({\parv})=\{\nabla v_{q+1}({\parv}),\ldots, \nabla v_T({\parv})\}^\top$. It can then be proved that
\begin{equation*}
    DQ_{IS}=\frac{\hbHit^\top \bX(\hparv)(\hbM_T \hbM^\top_T)^{-1}  \bX^\top(\hparv)\hbHit}{\alpha\,(1-\alpha)}\overset{d}{\to}\chi^2_q,
\end{equation*}
that in the reminder will be denoted as the \emph{In-Sample Dynamic Quantile} test statistic.

\subsection{Out-of-sample diagnostic tests for \VaR\ forecasts: the out-of-sample Dynamic Quantile test \citep{caviar}}
\label{ss:dvar}

Formally, let $(\valfor{T+1},\ldots,\valfor{T+H})$ be a sequence of ($1$-step ahead) out-of-sample \VaR\ forecasts. The OOS-DQ test statistic is computed as
\begin{equation}
	DQ_{OOS}=\frac{\tbHit(\estiv)^\top \tbX(\estiv)\{\tbX(\estiv)^\top\tbX(\estiv)\}^{-1}\tbX(\estiv)^\top \tbHit(\estiv)^\top}{H\alpha(1-\alpha)},
\end{equation}
where $\parv\subseteq\btheta$ indicates the subvector of model parameters involved in the \VaR\ model and
\begin{equation*}
    \tbHit(\estiv)=(\tHit_{T+q+1},\ldots,\tHit_{T+H}),
\end{equation*}
is the series of out-of-sample hits based on parameters estimated relying on information up to time $T$; $\tbX(\estiv)$ is a $(H-q)\times (q+2)$ matrix such that its $(t-q)$-th row is given by 
\begin{equation*}
    \tbX_t(\estiv)=(1, \tHit_{t-1},\ldots, \tHit_{t-q},\val{t}(\estiv)),
\end{equation*}
for $t=q+1,\ldots, H$, and where $\estiv$ is the estimate of $\parv$. 
It can be shown that, under the assumption in \cite{caviar}, $DQ_{OOS}\overset{d}{\to} \chi^2_{q+2}$.

The out-of-sample DQ test could also be implemented by augmenting the $\tbX(\hbtheta_T)$ with other instruments such as past volatility measures, such as realized variances, squared or absolute returns. In this case, the degrees of freedom of the $\chi^2$ distribution should be changed accordingly. 

The out-of-sample DQ test, in the above-presented configuration, can be seen as a portmanteau test for the correct specification of the \VaR\ forecasting model and, in this sense, we will refer to this test as the correct conditional coverage $DQ_{OOS}$ test, abbreviated $CC-DQ_{OOS}$. Differently, removing the constant term from the specification of $\tbX(\hbtheta_T)$ would yield a different version of the out-of-sample DQ test, that we will call the independence $DQ_{OOS}$ or, abbreviated, the $ID-DQ_{OOS}$. The name derives from the fact that $ID-DQ_{OOS}$ can detect serial correlation in the sequence of hits but, due to the missing constant term, cannot be used to assess correct coverage of \VaR\ forecasts. The asymptotic distribution for the $ID-DQ_{OOS}$ will be given by a $\chi^2_{q+1}$ random variable. Similarly, it is possible to define analogous conditional coverage and independence versions of the in-sample DQ test. In the remainder, these will be labeled as $CC-DQ_{IS}$ and $ID-DQ_{IS}$, respectively.

\subsection{Diagnostic tests for \ES\ forecasts}
\label{s:des}
\subsection{The ESR backtests \citep{bay_dim_2020}}
In \cite{bay_dim_2020}, authors propose a set of backtesting procedures for \ES\ regression (ESR), which build upon the testing approach introduced by \cite{minzar1969}. Specifically, the authors propose three ESR backtests: the ``auxiliary'', ``strict'' and ``strict intercept'' ESR.

The ``auxiliary'' ESR test is based on the bivariate regression model
\begin{eqnarray}
	\label{e:bd1}
    r_t&=&\beta_0+\beta_1 \valfor{t}+u^{v}_t,\\
    \label{e:bd2}
    r_t&=&\gamma_0+\gamma_1 \esfor{t}+u^{e}_t,
\end{eqnarray}
for $t = 1,\ldots,T$, where $\E(u^e_t|\mathcal{I}_{t-1},r_t<\valfor{t})=0$ and $Q_\alpha(u^v_t|\mathcal{I}_{t-1}) = 0$. The idea is that a series of \ES\ forecasts from a correctly specified \ES\ model should satisfy the following relation
\begin{equation*}
    \E(r_t|\mathcal{I}_{t-1}, r_t<\valfor{t})=\gamma_0+\gamma_1 \esfor{t},
\end{equation*}
with $(\gamma_0,\gamma_1)^\top=(0,1)^\top$. This hypothesis can be tested by fitting the regression model in (\ref{e:bd1}-\ref{e:bd2}) through the minimization of a strictly consistent joint (\VaR,\ \ES) loss function. This leads to the following Wald-type test statistic
\begin{equation*}
    T_{A-ESR}=T(\hbgamma-{\bgamma}_0)\widehat{\Omega}^{-1}_{\bgamma}(\hbgamma-{\bgamma}_0)^\top\overset{d}{\to}\chi^2_2,
\end{equation*}
where ${\bgamma}_0=(0,1)^\top$, $\hbgamma$ is a consistent estimator of $\bgamma=(\gamma_0,\gamma_1)$ and  $\widehat{\Omega}_{\bgamma}$ is a consistent estimator of the covariance of $\hbgamma$ \citep[see][]{bay_dim_2020}. The "strict" ESR test is based on a similar framework but $\valfor{t}$ in equation (\ref{e:bd1}) is replaced by $\esfor{t}$. Finally, the "strict intercept" test replaces equation (\ref{e:bd2}) by the following
\begin{equation*}
    r_t - \hat{e}_t = \gamma_1 + u^{e}_t.
\end{equation*}
The null is now given by ${\gamma}_1=0$ against a one-sided or a two-sided alternative \footnote{Differently from the "strict" and "auxiliary" tests for which only a two-sided alternative was allowed.}. The test is performed by computing a standard $t$-type statistic based on the estimated asymptotic variance of $\hat{\gamma}_1$. The one-sided alternative is of interest for regulatory and, in general, risk management purposes. In this paper, our interest is mainly in detecting deviations from the ideal situation of correct specification of the risk forecasting models. Hence, we will focus only on the situation where a two-sided alternative is considered. To implement the ESR backtests, we use the \verb"Esback" R package provided by the same authors \citep{esback2020} for our empirical analysis.

\subsection{Regression based calibration tests for VaR and ES \citep{pattonetal2019}}
\label{s:patton}
\cite{pattonetal2019} propose OLS regression-based calibration tests for assessing the quality of \VaR\ and \ES\ forecasts. In the auxiliary regression equations used for implementing the tests, the dependent variables are given by the standardized generalized residuals
\begin{equation*}
	\lambda^{s}_{v,t}=I(r_t\leq \valfor{t})-\alpha\qquad
	 \lambda^{s}_{e,t}=\frac{1}{\alpha}I(r_t\leq \hat{v}_t)\frac{r_t}{\hat{e}_t}-1
\end{equation*}
for \VaR\ and \ES, respectively. Both $\lambda^{s}_{v,t}$ and $\lambda^{s}_{e,t}$  are conditionally zero mean under correct specification of the \VaR\ and \ES\ models$\E(\lambda^{s}_{v,t}|\mathcal{I}_{t-1})=0$ and $\E(\lambda^{s}_{e,t}|\mathcal{I}_{t-1})=0$, for all $t$. It is also worth noting that $\lambda^{s}_{v,t}=Hit_t$, the hit variable already defined for DQ tests. 

The test procedures, henceforth denoted as the PZC tests, are based on fitting by OLS the following regression models
\begin{align}
	\label{e:pzc1}
	\lambda^s_{v,t}&=a_{0,v}+a_{1,v}\lambda^s_{v,t-1}+a_{2,v}\hat{v}_t+u^v_t\\
	\label{e:pzc2}
	\lambda^s_{e,t}&=a_{0,e}+a_{1,e}\lambda^s_{v,t-1}+a_{2,e}\hat{e}_t+u^v_t,
\end{align}
where, under correct specifications, we have $\mathbf{a}_v=(a_{0,v},a_{1,v},a_{2,v})^\top=\mathbf{0}$ and $\mathbf{a}_e= (a_{0,e},a_{1,e},a_{2,e})^\top=\mathbf{0}$. The statistics for testing these hypotheses are computed as
\begin{equation*}
    PZC_v= \hat{\mathbf{a}}^\top_{v}\widehat{\Omega}^{-1}_{{v}}\hat{\mathbf{a}}_{v}\overset{d}{\to}\chi^2_2,\qquad\qquad
    PZC_e= \hat{\mathbf{a}}^\top_{e}\widehat{\Omega}^{-1}_{{e}}\hat{\mathbf{a}}_{e}\overset{d}{\to}\chi^2_2,
\end{equation*}
where $\widehat{\Omega}_{{v}}$ ($\widehat{\Omega}_{{e}}$) is a consistent estimator of the asymptotic covariance matrix of $\hat{\mathbf{a}}_{v}$ ($\hat{\mathbf{a}}_{e}$). In our empirical analysis, following \cite{pattonetal2019}, to estimate ${\Omega}_{{v}}$ and ${\Omega}_{{e}}$ a Newey-West estimator with 20 lags is used.

\newpage
\section*{Appendix C: Tables}
\setcounter{figure}{0} \renewcommand{\thefigure}{C.\arabic{figure}}
\setcounter{table}{0} \renewcommand{\thetable}{C.\arabic{table}}
\setcounter{subsection}{0} \renewcommand{\thesubsection}{C.\arabic{subsection}}
\label{s:appFigTab}


\begin{table}[htp!]
\begin{small}
\caption{Calibration tests for \VaR\ and \ES\ models \cite{pattonetal2019}; in-sample DQ conditional coverage (CC$_{IS}$) and independence test (ID$_{IS}$); Wald tests for skewness and kurtosis coefficients in \VaR\ and \ES\ models; "Str.", "Aux.", "Str.I." ES Regression (ESR) calibration test (\cite{bay_dim_2020}): non-rejection frequencies at the 0.05 significance level (full sample).} 
\begin{tabular}{lcl|cc|cc|cc|ccc}
\multicolumn{3}{c|}{$\alpha = 0.01$} & \multicolumn{2}{c|}{In-sample DQ} & \multicolumn{2}{c|}{Wald test} & \multicolumn{2}{c|}{Calibration test} & \multicolumn{3}{c}{ESR calibration test} \\
\VaR & \ES & Loss & CC$_{IS}$ & ID$_{IS}$ & $a_i = 0$ & $b_i = 0$ & \VaR & \ES & "Str." & "Aux." & "Str.I." \\
\hline
\texttt{mlt_lev} & --- & EM  & 0.712 & 0.727 & 0.005 &  ---  & 0.617 &  ---  &  ---  &  ---  &  ---  \\
\texttt{mlt_skk} & --- & EM  & 0.701 & 0.690 & 0.002 &  ---  & 0.661 &  ---  &  ---  &  ---  &  ---  \\
\texttt{mlt_sim} & --- & EM  & 0.684 & 0.589 &  ---  &  ---  & 0.684 &  ---  &  ---  &  ---  &  ---  \\
\texttt{add_lev} & --- & EM  & 0.198 & 0.746 & 0.010 &  ---  & 0.594 &  ---  &  ---  &  ---  &  ---  \\
\texttt{add_skk} & --- & EM  & 0.193 & 0.679 & 0.004 &  ---  & 0.625 &  ---  &  ---  &  ---  &  ---  \\
\texttt{add_sim} & --- & EM  & 0.235 & 0.562 &  ---  &  ---  & 0.655 &  ---  &  ---  &  ---  &  ---  \\
\texttt{mlt_lev} & sim & ALS &  ---  &  ---  & 0.016 &  ---  & 0.548 & 0.695 & 0.988 & 0.987 & 0.995 \\
\texttt{mlt_skk} & sim & ALS &  ---  &  ---  & 0.017 &  ---  & 0.584 & 0.734 & 0.984 & 0.983 & 0.995 \\
\texttt{mlt_sim} & sim & ALS &  ---  &  ---  &  ---  &  ---  & 0.612 & 0.783 & 0.980 & 0.982 & 0.996 \\
\texttt{add_lev} & sim & ALS &  ---  &  ---  & 0.016 &  ---  & 0.563 & 0.741 & 0.984 & 0.984 & 0.994 \\
\texttt{add_skk} & sim & ALS &  ---  &  ---  & 0.010 &  ---  & 0.601 & 0.778 & 0.980 & 0.978 & 0.991 \\
\texttt{add_sim} & sim & ALS &  ---  &  ---  &  ---  &  ---  & 0.588 & 0.682 & 0.983 & 0.982 & 0.993 \\
\texttt{mlt_lev} & skk & ALS &  ---  &  ---  & 0.017 & 0.532 & 0.546 & 0.670 & 0.969 & 0.961 & 0.999 \\
\texttt{mlt_skk} & skk & ALS &  ---  &  ---  & 0.010 & 0.539 & 0.597 & 0.688 & 0.962 & 0.947 & 0.999 \\
\texttt{mlt_sim} & skk & ALS &  ---  &  ---  &  ---  & 0.494 & 0.638 & 0.718 & 0.978 & 0.931 & 0.999 \\
\texttt{add_lev} & skk & ALS &  ---  &  ---  & 0.012 & 0.610 & 0.598 & 0.722 & 0.975 & 0.946 & 0.997 \\
\texttt{add_skk} & skk & ALS &  ---  &  ---  & 0.015 & 0.589 & 0.604 & 0.730 & 0.975 & 0.946 & 0.997 \\
\texttt{add_sim} & skk & ALS &  ---  &  ---  &  ---  & 0.527 & 0.618 & 0.651 & 0.973 & 0.927 & 0.997 \\
\texttt{mlt_lev} & sim & FZ0 &  ---  &  ---  & 0.021 &  ---  & 0.554 & 0.716 & 0.982 & 0.985 & 0.995 \\
\texttt{mlt_skk} & sim & FZ0 &  ---  &  ---  & 0.021 &  ---  & 0.590 & 0.727 & 0.985 & 0.985 & 0.994 \\
\texttt{mlt_sim} & sim & FZ0 &  ---  &  ---  &  ---  &  ---  & 0.614 & 0.783 & 0.982 & 0.980 & 0.994 \\
\texttt{add_lev} & sim & FZ0 &  ---  &  ---  & 0.011 &  ---  & 0.560 & 0.737 & 0.984 & 0.983 & 0.994 \\
\texttt{add_skk} & sim & FZ0 &  ---  &  ---  & 0.019 &  ---  & 0.605 & 0.761 & 0.974 & 0.979 & 0.992 \\
\texttt{add_sim} & sim & FZ0 &  ---  &  ---  &  ---  &  ---  & 0.597 & 0.686 & 0.983 & 0.983 & 0.993 \\
\texttt{mlt_lev} & skk & FZ0 &  ---  &  ---  & 0.018 & 0.521 & 0.551 & 0.651 & 0.965 & 0.953 & 0.999 \\
\texttt{mlt_skk} & skk & FZ0 &  ---  &  ---  & 0.007 & 0.527 & 0.590 & 0.699 & 0.970 & 0.952 & 0.997 \\
\texttt{mlt_sim} & skk & FZ0 &  ---  &  ---  &  ---  & 0.487 & 0.623 & 0.711 & 0.967 & 0.934 & 0.999 \\
\texttt{add_lev} & skk & FZ0 &  ---  &  ---  & 0.019 & 0.597 & 0.593 & 0.732 & 0.970 & 0.950 & 0.999 \\
\texttt{add_skk} & skk & FZ0 &  ---  &  ---  & 0.010 & 0.592 & 0.618 & 0.738 & 0.968 & 0.942 & 0.997 \\
\texttt{add_sim} & skk & FZ0 &  ---  &  ---  &  ---  & 0.539 & 0.609 & 0.651 & 0.972 & 0.929 & 0.997 \\
\hline
\end{tabular}%
\label{tab:1}%
\end{small}
\end{table}%

\begin{table}[htp!]
\centering
\caption{Out-of-sample  DQ independence test (ID$_{OOS}$) and conditional coverage test (CC$_{OOS}$); calibration tests for \VaR\ and \ES\ models \cite{pattonetal2019}; "Strict", "Auxiliary" and "Strict Intercept" ES Regression (ESR) calibration test (\cite{bay_dim_2020}): non-rejection frequencies at the 0.05 significance level and number of valid assets  (out-of-sample data $T_{in}$).}
\begin{tabular}{c|lc|cc|cc|ccc}
& \multicolumn{2}{c|}{$\alpha = 0.01$} & \multicolumn{2}{c|}{Out-of-sample DQ} & \multicolumn{2}{c|}{Calibration test} & \multicolumn{3}{c}{ESR calibration test} \\
& \VaR & \ES & ID$_{OOS}$ & CC$_{OOS}$ & \VaR & \ES & "Str." & "Aux." & "Str.I."\\
\hline
\parbox[t]{2mm}{\multirow{12}{*}{\rotatebox[origin=c]{90}{$T_{in} = 500$}}} & \texttt{mlt_lev} & --- & 0.030 & 0.002 & 0.002 &  ---  &  ---  &  ---  &  ---  \\
&\texttt{mlt_skk} & --- & 0.010 & 0.000 & 0.005 &  ---  &  ---  &  ---  &  ---  \\
&\texttt{mlt_sim} & --- & 0.039 & 0.017 & 0.084 &  ---  &  ---  &  ---  &  ---  \\
&\texttt{add_lev} & --- & 0.025 & 0.000 & 0.002 &  ---  &  ---  &  ---  &  ---  \\
&\texttt{add_skk} & --- & 0.015 & 0.000 & 0.000 &  ---  &  ---  &  ---  &  ---  \\
&\texttt{add_sim} & --- & 0.052 & 0.022 & 0.074 &  ---  &  ---  &  ---  &  ---  \\
&\texttt{mlt_lev} & sim & 0.052 & 0.022 & 0.000 & 0.025 & 0.169 & 0.176 & 0.817 \\
&\texttt{mlt_skk} & sim & 0.020 & 0.007 & 0.007 & 0.057 & 0.186 & 0.211 & 0.838 \\
&\texttt{mlt_sim} & sim & 0.042 & 0.022 & 0.067 & 0.099 & 0.370 & 0.366 & 0.867 \\
&\texttt{add_lev} & sim & 0.022 & 0.005 & 0.000 & 0.074 & 0.266 & 0.261 & 0.884 \\
&\texttt{add_skk} & sim & 0.049 & 0.015 & 0.007 & 0.074 & 0.343 & 0.340 & 0.935 \\
&\texttt{add_sim} & sim & 0.015 & 0.007 & 0.049 & 0.092 & 0.317 & 0.348 & 0.898 \\\hline
\parbox[t]{2mm}{\multirow{12}{*}{\rotatebox[origin=c]{90}{$T_{in} = 1000$}}} & \texttt{mlt_lev} & --- & 0.180 & 0.086 & 0.165 &  ---  &  ---  &  ---  &  ---  \\
&\texttt{mlt_skk} & --- & 0.133 & 0.076 & 0.207 &  ---  &  ---  &  ---  &  ---  \\
&\texttt{mlt_sim} & --- & 0.224 & 0.185 & 0.405 &  ---  &  ---  &  ---  &  ---  \\
&\texttt{add_lev} & --- & 0.126 & 0.067 & 0.163 &  ---  &  ---  &  ---  &  ---  \\
&\texttt{add_skk} & --- & 0.126 & 0.047 & 0.175 &  ---  &  ---  &  ---  &  ---  \\
&\texttt{add_sim} & --- & 0.175 & 0.131 & 0.430 &  ---  &  ---  &  ---  &  ---  \\
&\texttt{mlt_lev} & sim & 0.143 & 0.089 & 0.064 & 0.114 & 0.409 & 0.415 & 0.948 \\
&\texttt{mlt_skk} & sim & 0.143 & 0.101 & 0.106 & 0.146 & 0.457 & 0.463 & 0.944 \\
&\texttt{mlt_sim} & sim & 0.190 & 0.170 & 0.326 & 0.304 & 0.641 & 0.688 & 0.964 \\
&\texttt{add_lev} & sim & 0.094 & 0.057 & 0.059 & 0.126 & 0.491 & 0.499 & 0.944 \\
&\texttt{add_skk} & sim & 0.126 & 0.069 & 0.099 & 0.168 & 0.487 & 0.487 & 0.939 \\
&\texttt{add_sim} & sim & 0.145 & 0.123 & 0.277 & 0.296 & 0.551 & 0.580 & 0.960 \\\hline
\parbox[t]{2mm}{\multirow{12}{*}{\rotatebox[origin=c]{90}{$T_{in} = 2000$}}} & \texttt{mlt_lev} & --- & 0.330 & 0.241 & 0.449 &  ---  &  ---  &  ---  &  ---  \\
&\texttt{mlt_skk} & --- & 0.310 & 0.227 & 0.491 &  ---  &  ---  &  ---  &  ---  \\
&\texttt{mlt_sim} & --- & 0.419 & 0.362 & 0.531 &  ---  &  ---  &  ---  &  ---  \\
&\texttt{add_lev} & --- & 0.281 & 0.200 & 0.459 &  ---  &  ---  &  ---  &  ---  \\
&\texttt{add_skk} & --- & 0.268 & 0.192 & 0.447 &  ---  &  ---  &  ---  &  ---  \\
&\texttt{add_sim} & --- & 0.377 & 0.335 & 0.578 &  ---  &  ---  &  ---  &  ---  \\
&\texttt{mlt_lev} & sim & 0.291 & 0.232 & 0.316 & 0.326 & 0.593 & 0.587 & 0.885 \\
&\texttt{mlt_skk} & sim & 0.234 & 0.192 & 0.358 & 0.356 & 0.629 & 0.643 & 0.886 \\
&\texttt{mlt_sim} & sim & 0.335 & 0.291 & 0.499 & 0.488 & 0.740 & 0.751 & 0.866 \\
&\texttt{add_lev} & sim & 0.303 & 0.207 & 0.294 & 0.311 & 0.656 & 0.644 & 0.880 \\
&\texttt{add_skk} & sim & 0.222 & 0.150 & 0.259 & 0.323 & 0.630 & 0.618 & 0.896 \\
&\texttt{add_sim} & sim & 0.300 & 0.241 & 0.486 & 0.472 & 0.735 & 0.735 & 0.872 \\
\hline
    \end{tabular}%
	\label{tab:2}%
\end{table}%

\begin{table}[htp!]
  \centering
  \caption{Median loss function values across all assets (out-of-sample data, $T_{in}=500,1000,2000$). For VaR models, the average quantile loss is reported ($\times 1000$). For ES models, the ALS scoring function is considered. The best model within each class is reported in boldface.}
    \begin{tabular}{ll|rrrrrrrrr}
          \VaR & \ES &  \multicolumn{3}{c}{$T_{in}=500$} & \multicolumn{3}{c}{$T_{in}=1000$} & \multicolumn{3}{c}{$T_{in}=2000$} \\
              \hline
          && 0.01  & 0.025 & 0.05  & 0.01  & 0.025 & 0.05  & 0.01  & 0.025 & 0.05 \\
              \hline
    \texttt{mlt_lev} & ---  & 0.999 & 1.667 & 2.538 & 0.852 & 1.490 & 2.322 & 0.818 & 1.464 & 2.319 \\
    \texttt{mlt_skk} & ---  & 0.982 & 1.643 & 2.521 & 0.842 & 1.486 & 2.309 & 0.801 & 1.458 & 2.320 \\
    \texttt{mlt_sim} & ---  & \textbf{0.940} & \textbf{1.602} & \textbf{2.454} & 0.822 & \textbf{1.469} & \textbf{2.291} & \textbf{0.790} & \textbf{1.443} & \textbf{2.308} \\
    \texttt{add_lev} & ---  & 0.990 & 1.663 & 2.524 & 0.855 & 1.494 & 2.330 & 0.813 & 1.465 & 2.322 \\
    \texttt{add_skk} & ---  & 0.992 & 1.648 & 2.537 & 0.859 & 1.490 & 2.320 & 0.816 & 1.467 & 2.325 \\
    \texttt{add_sim} & ---  & 0.958 & 1.639 & 2.569 & \textbf{0.820} & 1.472 & 2.306 & 0.798 & 1.448 & 2.315 \\
        \hline

    \texttt{mlt_lev} & sim  & -0.411 & -1.312 & -1.739 & -1.031 & -1.654 & -1.977 & -1.395 & -1.789 & -2.042 \\
    \texttt{mlt_skk} & sim  & -0.484 & -1.365 & -1.765 & -1.082 & -1.645 & -1.984 & -1.399 & -1.797 & -2.043 \\
    \texttt{mlt_sim} & sim  & \textbf{-1.002} & \textbf{-1.592} & -1.888 & \textbf{-1.319} & \textbf{-1.784} & \textbf{-2.045} & \textbf{-1.506} & \textbf{-1.844} & -2.055 \\
    \texttt{add_lev} & sim  & -0.354 & -1.316 & -1.737 & -1.014 & -1.651 & -1.965 & -1.357 & -1.762 & -2.026 \\
    \texttt{add_skk} & sim  & -0.442 & -1.331 & -1.695 & -0.968 & -1.604 & -1.954 & -1.297 & -1.756 & -2.023 \\
    \texttt{add_sim} & sim  & -0.665 & -1.568 & \textbf{-1.900} & -1.302 & -1.782 & -2.043 & -1.491 & -1.836 & \textbf{-2.064} \\
        \hline

    \end{tabular}%
  \label{tab:medianloss}%
\end{table}%

\begin{table}[htp!]
  \centering
   \caption{Average ranks across all assets based on quantile loss and ALS scoring functions (out-of-sample data, $T_{in}=500,1000,2000$).The best model within each class is reported in boldface.}
    \begin{tabular}{lc|rrrrrrrrr}
          \VaR & \ES &  \multicolumn{3}{c}{$T_{in}=500$} & \multicolumn{3}{c}{$T_{in}=1000$} & \multicolumn{3}{c}{$T_{in}=2000$} \\
              \hline
          && 0.01  & 0.025 & 0.05  & 0.01  & 0.025 & 0.05  & 0.01  & 0.025 & 0.05 \\
              \hline
    \texttt{mlt_lev} & --- & 4.441 & 4.374 & 4.153 & 4.303 & 4.241 & 4.155 & 4.204 & 4.034 & 3.975 \\
    \texttt{mlt_skk} & --- & 3.436 & 3.303 & 3.192 & 3.554 & 3.594 & 3.483 & 3.643 & 3.466 & 3.581 \\
    \texttt{mlt_sim} & --- & \textbf{2.002} & \textbf{1.709} & \textbf{1.820 }& \textbf{2.020} & \textbf{2.027} & \textbf{2.081} & \textbf{2.261} &\textbf{2.241} & \textbf{2.264} \\
    \texttt{add_lev} & --- & 4.209 & 4.562 & 4.569 & 4.475 & 4.781 & 4.751 & 4.387 & 4.562 & 4.369 \\
    \texttt{add_skk} & --- & 3.990 & 4.143 & 4.081 & 4.143 & 4.113 & 4.217 & 4.165 & 4.236 & 4.200 \\
    \texttt{add_sim} & --- & 2.921 & 2.909 & 3.185 & 2.505 & 2.244 & 2.313 & 2.340 & 2.461 & 2.611 \\
    \hline
    \texttt{mlt_lev} & sim  & 3.956 & 4.160 & 4.025 & 4.076 & 4.030 & 4.084 & 4.002 & 4.002 & 3.803 \\
    \texttt{mlt_skk} & sim  & 3.739 & 3.685 & 3.749 & 3.766 & 3.938 & 3.778 & 3.778 & 3.682 & 3.734 \\
    \texttt{mlt_sim} & sim  & \textbf{2.067} & \textbf{2.229} & \textbf{2.377} & \textbf{2.057} & \textbf{2.167} & 2.246 & \textbf{2.094} & \textbf{2.131} & \textbf{2.091} \\
    \texttt{add_lev} & sim  & 4.101 & 4.283 & 4.165 & 4.234 & 4.458 & 4.507 & 4.446 & 4.549 & 4.623 \\
    \texttt{add_skk} & sim  & 3.823 & 4.126 & 4.227 & 4.251 & 4.350 & 4.406 & 4.515 & 4.505 & 4.552 \\
    \texttt{add_sim} & sim  & 3.315 & 2.517 & 2.458 & 2.616 & 2.057 & \textbf{1.978} & 2.165 & \textbf{2.131} & 2.197 \\
        \hline

    \end{tabular}%
  \label{tab:medianranks}%
\end{table}%

\newpage
\section*{Appendix D: Figures}
\setcounter{figure}{0} \renewcommand{\thefigure}{D.\arabic{figure}}
\setcounter{table}{0} \renewcommand{\thetable}{D.\arabic{table}}
\setcounter{subsection}{0} \renewcommand{\thesubsection}{D.\arabic{subsection}}
\label{s:appFigTab}

\begin{figure}[htp!]
	\begin{center}
	\includegraphics[height=0.87\textheight]{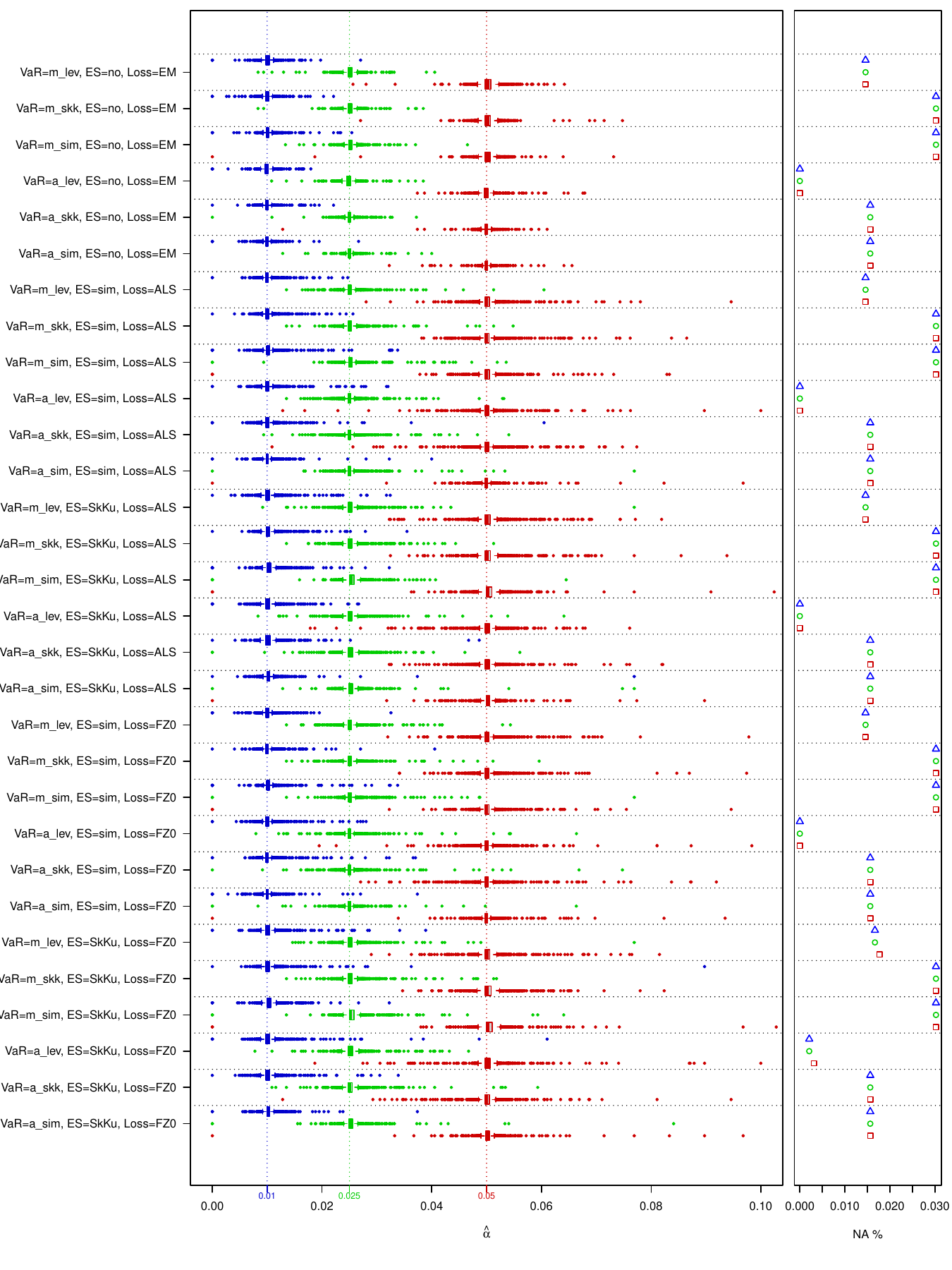}
\caption{In-sample coverage for the models listed on the Y-axis. The left panel displays the estimated coverage probabilities (\(\hat{\alpha}\)), while the right panel shows the percentage of cases where the model estimation failed. Different colors represent different true coverage levels: blue for \(\alpha = 0.01\), green for \(\alpha = 0.025\), and red for \(\alpha = 0.05\). Each point corresponds to one stock.}\label{fig:insampleCov}
\end{center}
\end{figure}
\begin{figure}[htp!]
	\begin{center}
	\includegraphics[height=0.92\textheight]{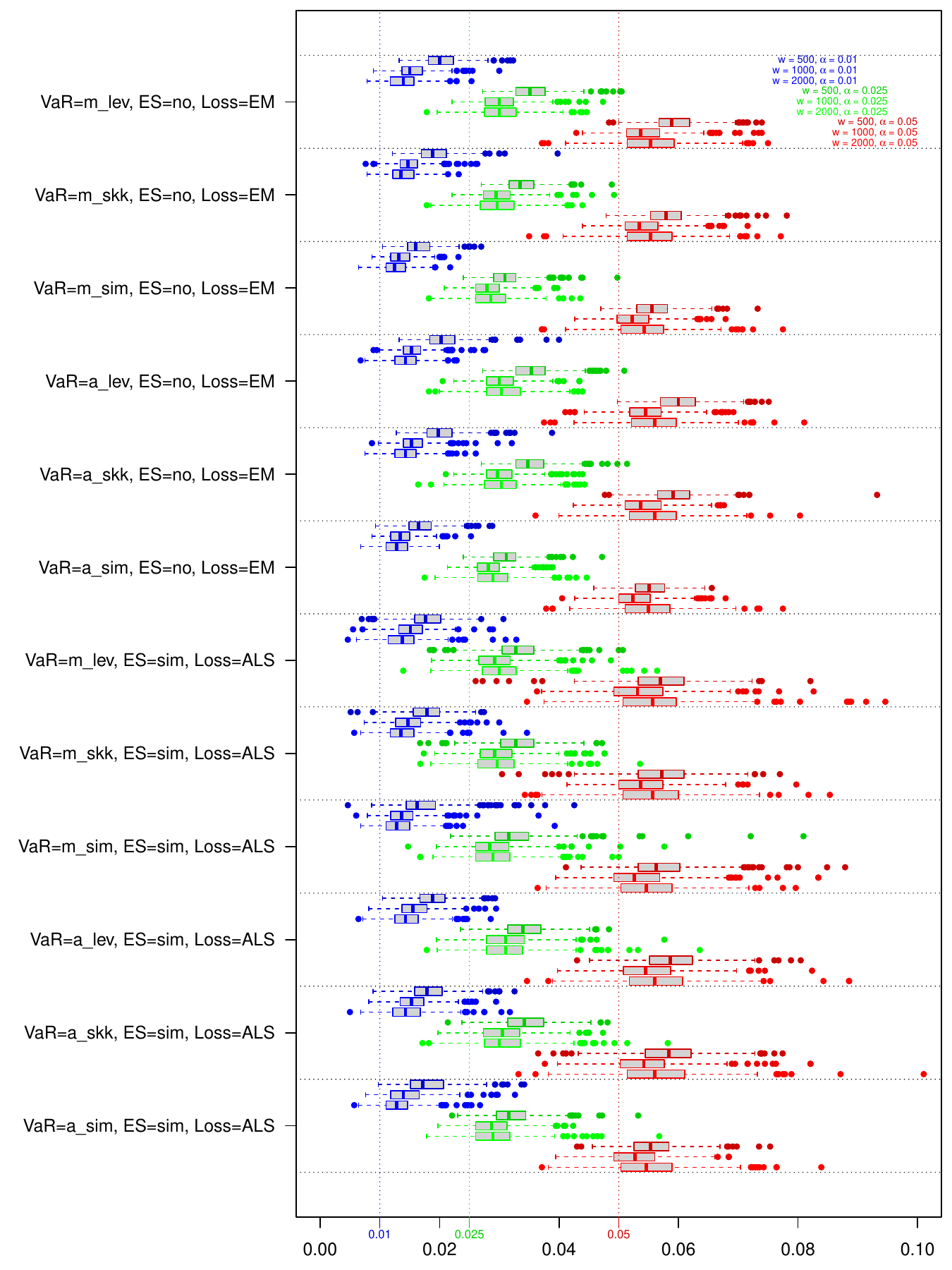}
    \caption{Out-sample-coverage estimated probabilities (\(\hat{\alpha}\)) for the models listed on the Y-axis. Different colors represent different true coverage levels: blue for \(\alpha = 0.01\), green for \(\alpha = 0.025\), and red for \(\alpha = 0.05\). Different intensities of the colours represent different widths of the rolling window, as depicted in top-right corner. Each point corresponds to one stock.}\label{fig:outofsampleCov}
\end{center}
\end{figure}
\begin{figure}[htp!]
	\begin{center}
	\includegraphics[height=0.92\textheight]{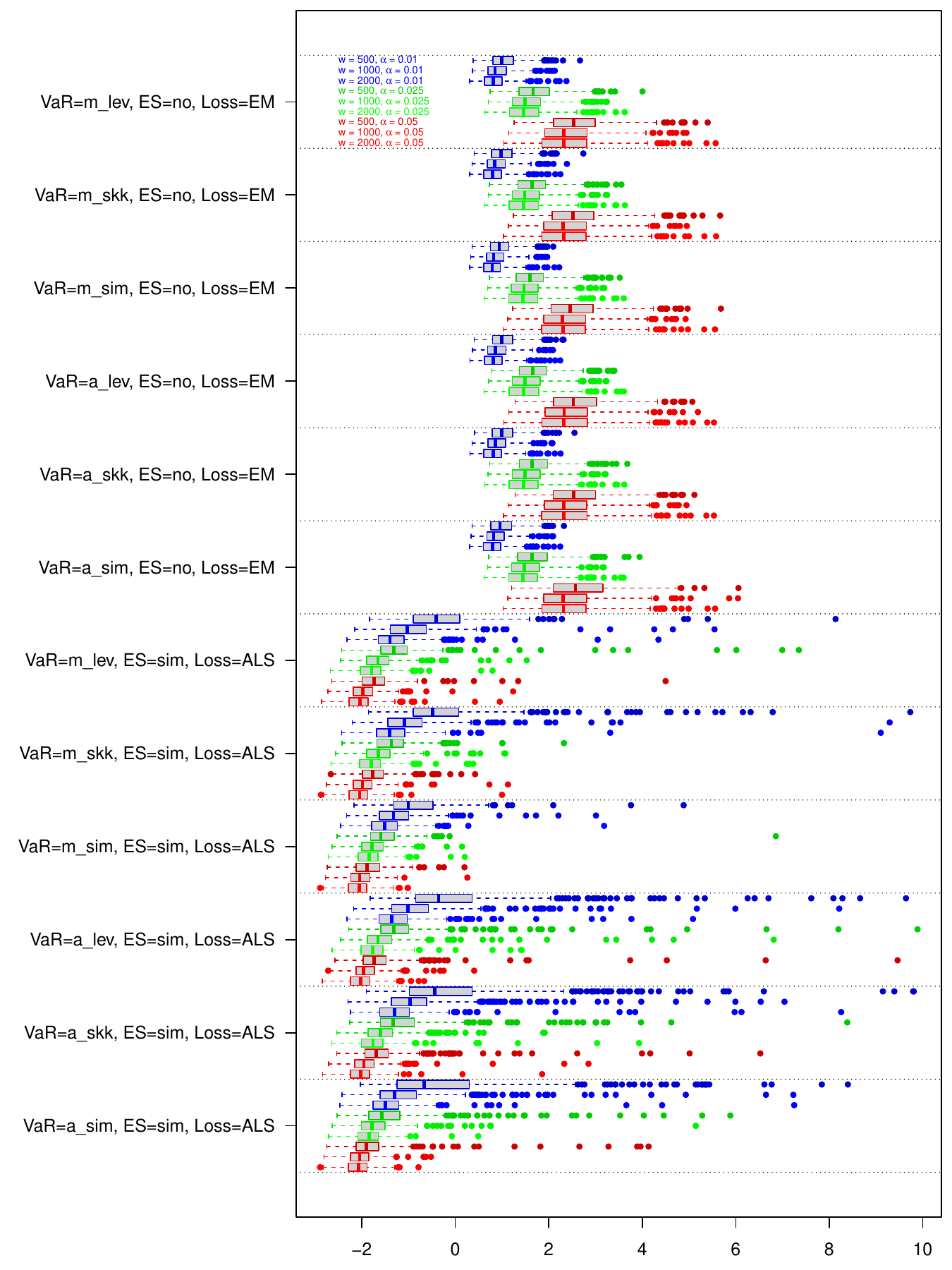}
    \caption{Value of the loss function in out-sample-performance for the models listed on the Y-axis. Different colors represent different true coverage levels: blue for \(\alpha = 0.01\), green for \(\alpha = 0.025\), and red for \(\alpha = 0.05\). Different intensities of the colours represent different widths of the rolling window, as depicted in top-right corner. Each point corresponds to one stock.}\label{fig:outofsampleLosses}
    \end{center}
\end{figure}

\end{document}